\documentclass[12pt]{article}
\usepackage{epsf}

\newcommand{\dlt}{\delta}
\newcommand{\Dlt}{\Delta}
\newcommand{\ep}{\epsilon}
\newcommand{\gm}{\gamma}

\newcommand{\tht}{\theta}
\newcommand{\btht}{\bar{\tht}}
\newcommand{\lmd}{\lambda}
\newcommand{\Lmd}{\Lambda}
\newcommand{\sgm}{\sigma}
\newcommand{\vph}{\varphi}

\newcommand{\be}{\begin{equation}}
\newcommand{\ee}{\end{equation}}
\newcommand{\bea}{\begin{eqnarray}}
\newcommand{\eea}{\end{eqnarray}}
\newcommand{\eql}{\!\!\!&=\!\!\!&}

\newcommand{\defa}{\!\!\!&\equiv\!\!\!&}

\newcommand{\mtrx}[4]{\brkt{\begin{array}{cc}#1&#2\\#3&#4\end{array}}}
\newcommand{\vct}[2]{\brkt{\begin{array}{c}#1\\#2\end{array}}}

\newcommand{\tl}[1]{\tilde{#1}}

\newcommand{\diag}{{\rm diag}}
\newcommand{\der}{\partial}
\newcommand{\dr}{\!\!{\rm d}}

\newcommand{\Acl}{A_{\rm cl}}
\newcommand{\Acla}{A_{\rm cl1}}
\newcommand{\Aclb}{A_{\rm cl2}}
\newcommand{\AI}{A_{\rm I}}
\newcommand{\AII}{A_{\rm II}}
\newcommand{\vpcl}{\varphi_{\rm cl}}
\newcommand{\uR}[1]{u_{{\rm R}(#1)}}
\newcommand{\uI}[1]{u_{{\rm I}(#1)}}
\newcommand{\UR}{U_{{\rm R}0}}
\newcommand{\UI}{U_{{\rm I}0}}
\newcommand{\psR}[1]{\psi_{{\rm R}(#1)}}
\newcommand{\psI}[1]{\psi_{{\rm I}(#1)}}
\newcommand{\uRX}[1]{u_{{\rm R}X(#1)}}
\newcommand{\uIX}[1]{u_{{\rm I}X(#1)}}
\newcommand{\psRX}[1]{\psi_{{\rm R}X(#1)}}
\newcommand{\psIX}[1]{\psi_{{\rm I}X(#1)}}
\newcommand{\befa}{\beta_{{\rm eff}1}}
\newcommand{\befb}{\beta_{{\rm eff}2}}
\newcommand{\cn}{{\rm cn}}
\newcommand{\dn}{{\rm dn}}

\newcommand{\brkt}[1]{\left( #1 \right)}
\newcommand{\brc}[1]{\left\{ #1 \right\}}
\newcommand{\sbk}[1]{\left[ #1 \right]}

\renewcommand{\Re}{{\rm Re}}
\renewcommand{\Im}{{\rm Im}}

\newcommand{\cA}{{\cal A}}

\newcommand{\cL}{{\cal L}}

\newcommand{\cN}{{\cal N}}

\newcommand{\NP}[1]{{\it Nucl.~Phys.}~{\bf #1}}
\newcommand{\PL}[1]{{\it Phys.~Lett.}~{\bf #1}}

\newcommand{\MPL}[1]{{\it Mod.~Phys.~Lett.}~{\bf #1}}

\newcommand{\PR}[1]{{\it Phys.~Rev.}~{\bf #1}}
\newcommand{\PRL}[1]{{\it Phys.~Rev.~Lett.}~{\bf #1}}

\newcommand{\JHEP}[1]{{\it JHEP}~{\bf #1}}

\begin{document}

\begin{titlepage}
\null
\begin{flushright}
 {\tt hep-th/0304195}\\
TU-688
\\
Apr 2003
\end{flushright}

\vskip 2cm
\begin{center}
{\LARGE \bf  Effective theory for wall-antiwall system}

\lineskip .75em
\vskip 2.5cm

\normalsize

{\large \bf Yutaka Sakamura}
{\def\thefootnote{\fnsymbol{footnote}}
\footnote[5]{\it  e-mail address:
sakamura@tuhep.phys.tohoku.ac.jp}}

\vskip 1.5em

{\it Department of Physics, Tohoku University\\ 
Sendai 980-8578, Japan}

\vspace{18mm}

{\bf Abstract}\\[5mm]
{\parbox{13cm}{\hspace{5mm} \small
We propose a useful method for deriving the effective theory 
for a system where BPS and anti-BPS domain walls coexist. 
Our method respects an approximately preserved SUSY near each wall. 
Due to the finite width of the walls, SUSY breaking terms arise at tree-level, 
which are exponentially suppressed. 
A practical approximation using the BPS wall solutions is also discussed. 
We show that a tachyonic mode appears in the matter sector 
if the corresponding mode function has a broader profile 
than the wall width. 
}}

\end{center}

\end{titlepage}

\clearpage

\section{Introduction}
{\it Supersymmetry} (SUSY) is one of the most promising candidate 
for the solution to the hierarchy problem. 
When you construct a realistic model, however, SUSY must be broken 
because no superparticles have been observed yet. 
Thus, considering some SUSY breaking mechanism is an important issue 
for the realistic model-building. 
However, conventional SUSY breaking mechanisms are somewhat complicated 
in four dimensions \cite{ADS}. 

On the other hand, the {\it brane-world scenario}, which was originally 
proposed as an alternative solution to the hierarchy problem \cite{ADD,RS}, 
has been attracted much attention and investigated in various frameworks. 
It is based on an idea that our four-dimensional (4D) world is embedded 
into a higher dimensional space-time\footnote{The original idea along 
this direction was proposed in Ref.\cite{akama}. 
}. 
The 4D hypersurface on which we live is generally called a {\it brane}, 
and it can be a D brane (a stack of D branes) in the string theory 
or a topological defect in the field theory like a domain wall, 
vortex, monopole and so on. 
Recently, in spite of its original motivation of solving 
the hierarchy problem, the extra dimensions are utilized 
in order to explain various phenomenological problems, 
such as the fermion mass hierarchy \cite{arkani,DS2,kaplan}, 
the proton stability \cite{arkani}, 
the doublet-triplet splitting in the grand unified theory \cite{maru} 
and so on. 
In particular, the brane-world scenario can provide some simple mechanisms 
of SUSY breaking. 
One of them is a mechanism where SUSY is broken by the boundary conditions 
for the bulk fields \cite{AMQ}.\footnote{
Its basic idea was first proposed in Ref.\cite{scherk}. }

\begin{figure}[t]
 \leavevmode
 \epsfysize=5cm
 \centerline{\epsfbox{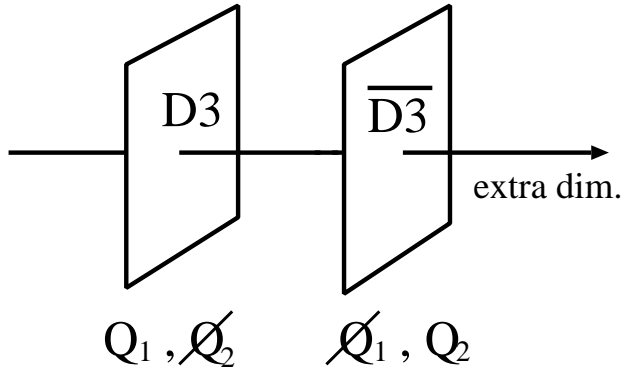}}
 \caption{The brane configuration of the pseudo-supersymmetry.}
 \label{pseudoSUSY}
\end{figure}

There is another interesting SUSY breaking mechanism, 
which utilizes {\it BPS objects}. 
They are the states that saturate the Bogomol'nyi-Prasad-Sommerfield (BPS) 
bound \cite{BPS}, and preserve part of the bulk SUSY. 
By using the (BPS) D branes, for example, we can break 
SUSY in a very simple way. 
This mechanism is called {\it pseudo-supersymmetry}~\cite{klein} or 
{\it quasi-supersymmetry}~\cite{cremades}. 
In this mechanism, the theory has two or more sectors 
in which different fractions of the bulk SUSY are preserved, 
and all of it are broken in the whole system. 
One of the simplest examples is a system where parallel D3 brane and 
anti-D3 brane coexist with a finite distance. 
(See Fig.\ref{pseudoSUSY}.) 
Since each brane preserves an opposite half of the bulk SUSY, 
SUSY is completely broken in this system. 
A 4D effective theory for such a system can be derived by using 
the technique of the nonlinear realization for the space-time 
symmetries~\cite{klein}. 

In the field theoretical context, 
there is a similar SUSY breaking mechanism to the pseudo-supersymmetry. 
In this mechanism, SUSY is broken due to the coexistence of 
two {\it BPS domain walls} which preserve 
opposite halves of the bulk SUSY \cite{DS2,MSSS2,MSSS,EMSS}. 
Therefore, the corresponding field configuration is non-BPS. 
We will refer to such a non-BPS configuration 
as a {\it wall-antiwall configuration} in this paper. 
One of the main differences from the pseudo-supersymmetry is 
that the BPS walls have a finite width unlike the D-branes. 
Due to this wall width, SUSY breaking terms arise at tree-level 
in contrast to the pseudo-supersymmetry scenario 
where they are induced by quantum effects. 
In general, such a non-BPS configuration is unstable \cite{MSSS2}. 
However, we can construct a stable wall-antiwall configuration 
by introducing a topological winding around the target space 
of the scalar fields \cite{MSSS}. 

When you discuss the phenomenological arguments on the wall-antiwall 
background, we need to derive a low-energy effective theory on it. 
If we eliminate the auxiliary fields of the superfields at the first step, 
the procedure of integrating out the massive modes becomes very complicated 
\cite{HLW}. 
In the case of the BPS-wall background, this complication can be avoided 
by keeping a half of the bulk SUSY preserved by the wall to be manifest, 
{\it i.e.}~by expanding the bulk superfields directly 
into lower-dimensional superfields 
for the unbroken SUSY \cite{sakamura1}.  
In the case of the wall-antiwall background, however, such a way out 
cannot be applied because there is no SUSY preserved in the whole system. 
Therefore, it is difficult to derive the effective theory 
on such a background correctly. 
In Ref.\cite{MSSS}, only mass-splitting between boson and fermion in 
each supermultiplet is calculated 
with the help of the low-energy theorem~\cite{lee,clark}.\footnote{
Some other SUSY breaking terms were also estimated in Ref.\cite{MSSS}, 
but they were only part of the whole terms.}

In this paper, we will propose a useful method for 
deriving the effective theory in the wall-antiwall system. 
To avoid inessential complications, we will consider three-dimensional 
(3D) walls in a 4D bulk theory throughout this paper. 
Our method respects the approximately supersymmetric structure 
near the walls, and the resultant Lagrangian is written by 
the 3D superfields and the SUSY breaking terms. 
We can calculate {\it all SUSY breaking parameters} systematically 
in our method. 
Using this method, we will show that a tachyonic mode appears 
in the effective theory from the matter sector 
if the corresponding mode function has a broader profile 
than the wall width. 
We will also provide an approximation method using the BPS wall solutions. 
This approximation is useful as a practical method
because the BPS wall solutions can easily be found than the non-BPS ones. 

In the superstring theory, it is well-known that 
BPS D-branes can be described by the kink solutions 
in the field theory of the tachyon \cite{MZ}. 
In particular, a D$p$ brane and anti-D$p$ brane pair is described 
as a wall-antiwall solution 
in an effective theory on a non-BPS D$(p+1)$ brane \cite{hashimoto}. 
Thus, our method can also be used 
in the discussion of the tachyon condensation \cite{tachyon} 
in the superstring theory. 

This paper is organized as follows. 
In the next section, we will provide useful expressions 
of the 4D bulk action, which respect opposite halves of the bulk SUSY. 
In Sect.\ref{nonBPSsys}, we will consider the wall-antiwall configuration 
and derive the 3D effective theory. 
A practical approximation method is provided in Sect.\ref{BPSapx}.  
In Sect.\ref{exmpl_model}, we will demonstrate our method 
in an explicit example of the stable wall-antiwall configuration.  
We will also discuss the gauge interactions in Sect.\ref{gauge_mpt}. 
Sect.\ref{summary} is devoted to the summary and the discussion. 
Notations and some useful properties of the classical background 
are listed in the appendices. 


\section{Useful expressions of the bulk action} \label{BPSformulae}
We will consider the following Wess-Zumino model as a 4D $\cN=1$ bulk theory. 
\be
 S=\int\dr^4 x{\rm d}^2\tht{\rm d}^2\btht\: \bar{\Phi}\Phi
 +\int\dr^4 x{\rm d}^2\tht\: W(\Phi)+\int\dr^4 x{\rm d}^2\btht\: \bar{W}(\bar{\Phi}), 
 \label{WZmodel}
\ee
where 
\be
 \Phi(y^\mu,\tht)=A(y^\mu)+\sqrt{2}\tht\Psi(y^\mu)+\tht^2F(y^\mu) \;\;\;\;\;
 (y^\mu\equiv x^\mu+i\tht\sgm^\mu\btht) 
\ee
is a chiral superfield. 

Here, we decompose the Grassmann coordinates $\tht$ and $\btht$ 
as\footnote{
The notations of $\tht_1$ and $\tht_2$ are different from those 
in Refs.\cite{sakamura1}.} 
\be
 \tht^\alpha = \frac{e^{i\dlt/2}}{\sqrt{2}}(\tht_2^\alpha+i\tht_1^\alpha), \;\;\;
 \btht^{\dot{\alpha}} = \frac{e^{-i\dlt/2}}{\sqrt{2}}(\tht_2^\alpha-i\tht_1^\alpha), 
 \label{tht_decomp}
\ee
where $\dlt$ is a constant phase. 
Corresponding to this decomposition, we also decompose the 
supercharges~$Q_\alpha$ and $\bar{Q}_{\dot{\alpha}}$ as 
\be
 Q_\alpha = \frac{e^{-i\dlt/2}}{\sqrt{2}}(Q_{2\alpha}-iQ_{1\alpha}), \;\;\;
 \bar{Q}_{\dot{\alpha}} = -(Q_\alpha)^* = -\frac{e^{i\dlt/2}}{\sqrt{2}}
 (Q_{2\alpha}+iQ_{1\alpha}).  \label{Q_decomp}
\ee
Then, it follows that 
\be
 \tht Q+\btht\bar{Q}=\tht_1Q_1+\tht_2Q_2. 
\ee
Under the above decomposition, the 4D $\cN=1$ SUSY algebra can be rewritten as 
\bea
 \brc{Q_{1\alpha},Q_{1\beta}}\eql \brc{Q_{2\alpha},Q_{2\beta}}
 =2\brkt{\gm_{(3)}^m\sgm^2}_{\alpha\beta}P_m, \nonumber\\
 \brc{Q_{1\alpha},Q_{2\beta}}\eql -\brc{Q_{2\alpha},Q_{1\beta}}
 =-2i\brkt{\sgm^2}_{\alpha\beta}P_2. 
\eea
This can be interpreted as the central extended 3D $\cN=2$ SUSY algebra 
if we identify $P_2$ with the central charge. 

Now, if we perform an integration for $\tht_2$ in Eq.(\ref{WZmodel}), 
we can obtain the following expression \cite{sakamura1}. 
\be
 S=\int\dr^3x{\rm d}^2\tht_1{\rm d}x_2 \: 
 \brc{D_1^\alpha\bar{\vph}^{(1)}D_{1\alpha}\vph^{(1)}
 +\Im\brkt{2\bar{\vph}^{(1)}\der_2\vph^{(1)}+4e^{-i\dlt}W(\vph^{(1)})}}, 
 \label{S_vph1}
\ee
where $\der_2\equiv\der/\der x_2$, 
$D_{1\alpha}\equiv \der/\der\tht_1^\alpha-i(\gm_{(3)}^m\tht_1)_\alpha\der_m$ 
is a covariant spinor derivative\footnote{
Throughout this paper, derivatives for Grassmann variables are regarded as 
left-derivatives.} for $Q_1$-SUSY, and 
\bea
 \vph^{(1)}(x^m,x_2,\tht_1)\defa e^{\tht_1Q_1}\times A(x^m,x_2) \nonumber\\
 \eql A+e^{i\dlt/2}(\sgm^2)_{\alpha\beta}\tht_1^\alpha\Psi^\beta
 +\frac{i}{2}\tht_1^2(\der_2 A+e^{i\dlt}F).  \label{def_vph1}
\eea
Throughout this paper, $m=0,1,3$ denotes the 3D Lorentz index, 
and the $x_2$-direction is chosen as the extra dimension. 
The relation between $\vph^{(1)}$ and the original chiral superfield~$\Phi$ 
is given by 
\be
 \Phi(y^\mu,\tht)=
 e^{-\tht_1\tht_2\der_2}e^{-i\tht_2D_1+i\tht_2^2\der_2}\vph^{(1)}
 (x^m,x_2,\tht_1).  \label{Phi_vph1}
\ee

On the other hand, if we perform an integration for $\tht_1$, Eq.(\ref{WZmodel}) 
becomes 
\be
 S=\int\dr^3x{\rm d}^2\tht_2{\rm d}x_2 \:
 \brc{D_2^\alpha\bar{\vph}^{(2)}D_{2\alpha}\vph^{(2)}
 +\Im\brkt{2\bar{\vph}^{(2)}\der_2\vph^{(2)}-4e^{-i\dlt}W(\vph^{(2)})}}, 
 \label{S_vph2}
\ee
where $D_{2\alpha}\equiv \der/\der\tht_2^\alpha-i(\gm_{(3)}^m\tht_2)_\alpha\der_m$ 
is a covariant spinor derivative for $Q_2$-SUSY, and 
\bea
 \vph^{(2)}(x^m,x_2,\tht_2)\defa e^{\tht_2Q_2}\times A(x^m,x_2) \nonumber\\
 \eql A-ie^{i\dlt/2}(\sgm^2)_{\alpha\beta}\tht_2^\alpha\Psi^\beta
 +\frac{i}{2}\tht_2^2(\der_2 A-e^{i\dlt}F). 
\eea
The relation between $\vph^{(2)}$ and $\Phi$ is given by 
\be
 \Phi(y^\mu,\tht)=
 e^{\tht_1\tht_2\der_2}e^{i\tht_1D_2+i\tht_1^2\der_2}\vph^{(2)}
 (x^m,x_2,\tht_2).  \label{Phi_vph2}
\ee

From Eqs.(\ref{Phi_vph1}) and (\ref{Phi_vph2}), we can obtain the relation 
between $\vph^{(1)}$ and $\vph^{(2)}$ as follows. 
\bea
 \vph^{(1)}(x^m,x_2,\tht_1) \eql \Phi|_{\tht_2=0} = e^{i\tht_1D_2+i\tht_1\der_2}
 \vph^{(2)}|_{\tht_2=0} \nonumber\\
 \eql \vph^{(2)}(x^m,x_2+i\tht_1^2,i\tht_1)  \label{rel_vph12}
\eea
In fact, we can explicitly show that 
\bea
 && \int\dr^2\tht_1{\rm d}x_2 \: 
 \left\{D_1^\alpha\brkt{\bar{\vph}^{(2)}(x^m,x_2-i\tht_1^2,-i\tht_1)}
 D_{1\alpha}\brkt{\vph^{(2)}(x^m,x_2+i\tht_1^2,i\tht_1)} \right. \nonumber\\
 && \hspace{2.5cm} \left.+\Im\brkt{2\bar{\vph}^{(2)}
 (x^m,x_2-i\tht_1^2,-i\tht_1)\der_2
 \brkt{\vph^{(2)}(x^m,x_2+i\tht_1^2,i\tht_1)}}\right\} \nonumber\\
 \eql \int\dr^2\tht_2{\rm d}x_2 \:
 \brc{D_2^\alpha\bar{\vph}^{(2)}D_{2\alpha}\vph^{(2)}
 +\Im\brkt{2\bar{\vph}^{(2)}\der_2\vph^{(2)}}}, 
 \label{K_vph1-vph2}
\eea
and 
\be
 \int\dr^2\tht_1{\rm d}x_2 \:
 \Im\brkt{e^{-i\dlt}W(\vph^{(2)}(x^m,x_2+i\tht_1^2,i\tht_1))} 
 = \int\dr^2\tht_2{\rm d}x_2 \:
 \Im\brkt{-e^{-i\dlt}W(\vph^{(2)})}. 
 \label{W_vph1-vph2}
\ee
Namely, we can derive Eq.(\ref{S_vph2}) by substituting Eq.(\ref{rel_vph12}) 
into Eq.(\ref{S_vph1}).

\section{Wall-antiwall system} \label{nonBPSsys}
\subsection{Classical field configuration}
Let us assume that the bulk theory~(\ref{WZmodel}) has two supersymmetric 
vacua $A=\AI$ and $A=\AII$, and that there exists a BPS domain wall 
configuration~$A=\Acla(x_2)$ which interpolates between them. 
That is, 
\be
 \Acla(-\infty)=\AI, \;\;\; \Acla(\infty)=\AII,
\ee
and $\Acla(x_2)$ satisfies a BPS equation 
\be
 \der_2 \Acla=e^{i\dlt}\bar{W}'(\bar{A}_{\rm cl1}),  \label{BPS_eq1}
\ee
where a prime denotes a differentiation for the argument, and 
\be
 \dlt\equiv \arg\brkt{W(\AII)-W(\AI)}. 
\ee
If we take this value of $\dlt$ as that in the decompositions~(\ref{tht_decomp}) 
and (\ref{Q_decomp}), $Q_1$ and $Q_2$ correspond to the unbroken and 
broken supercharges respectively. 

When the theory has the BPS domain wall~$\Acla(x_2)$, 
it also has the ``anti-BPS'' domain wall~$A=\Aclb(x_2)$. That is, 
\be
 \Aclb(-\infty)=\AII, \;\;\; \Aclb(\infty)=\AI, 
\ee
and the BPS equation for $\Aclb(x_2)$ is 
\be 
 \der_2 \Aclb=-e^{i\dlt}\bar{W}'(\bar{A}_{\rm cl2}).  \label{BPS_eq2}
\ee
This anti-BPS wall preserves $Q_2$-SUSY and breaks $Q_1$-SUSY. 

Now we will consider a system in which the above two kinds of BPS walls 
coexist. 
Since the two walls preserve opposite halves of the bulk SUSY, 
All of it are broken in such a system. 
That is, a corresponding field configuration~$\Acl(x_2)$ is non-BPS. 
Unlike the BPS wall, such a non-BPS configuration is usually unstable 
\cite{MSSS2}. 
However, we can construct a stable non-BPS configuration 
by introducing the topological winding around the target space of 
the scalar field~$A(x_2)$ \cite{MSSS}. 
We will discuss an explicit example of such a stable configuration 
in Sect.\ref{exmpl_model}. 
In the following, we will assume that the extra dimension ($x_2$-direction) 
is compactified on $S^1$ with radius~$R$. 

Furthermore, to simplify the discussions, we will also assume that 
the superpotential~$W$ has the following symmetric property\footnote{
This assumption is not essential for our method of deriving 
the effective theory. 
It is just a matter of convenience. 
}. 
\be
 W'(A-A_{\rm M})=W'(A_{\rm M}-A),  \label{W_assumption}
\ee
where $A_{\rm M}\equiv (\AI+\AII)/2$ is a middle point between 
the two vacua on the target space. 
Under this assumption, the BPS and anti-BPS walls are located 
at antipodal points on $S^1$, and $\Acl(x_2)$ has the following property. 
(See Fig.\ref{profile-Acl} in Sect.\ref{exmpl_model}.)
\bea
 \der_2\Acl(x_2-\pi R) \eql \pm\der_2\Acl(x_2), \nonumber\\
 W'(\Acl(x_2-\pi R)) \eql \mp W'(\Acl(x_2)).  \label{piR_translation}
\eea
The upper signs in R.H.S correspond to 
the case of a stable configuration with the topological winding \cite{MSSS}, 
and the lower signs correspond to the case of an unstable configuration 
with no winding \cite{MSSS2}. 
From now on, we will choose the origin of the coordinate~$x_2$ 
so that the (approximate) BPS wall is 
located at $x_2=0$ and the anti-BPS wall is at $x_2=\pi R$.

\subsection{Mode-expansion}
First, we will consider the mode-expansion of $\vph^{(1)}$ in Eq.(\ref{S_vph1}). 
The classical background of $\vph^{(1)}$ is 
\bea
 \vpcl^{(1)}(x_2,\tht_1) \eql e^{\tht_1Q_1}\times \Acl(x_2)
 =\Acl(x_2)+\frac{i}{2}\tht_1^2\brkt{\der_2\Acl(x_2)+e^{i\dlt}F_{\rm cl}(x_2)} 
 \nonumber\\
 \eql \Acl(x_2)+\frac{i}{2}\tht_1^2\brkt{\der_2\Acl(x_2)-e^{i\dlt}
 \bar{W}'(\bar{A}_{\rm cl}(x_2))}. 
\eea
Here we used the equation of motion for the auxiliary field~$F$ 
in the last equation. 

Similarly, the background of $\vph^{(2)}$ is 
\be
 \vpcl^{(2)}(x_2,\tht_2) = e^{\tht_2Q_2}\times \Acl(x_2) 
 =\Acl(x_2)+\frac{i}{2}\tht_2^2\brkt{\der_2\Acl(x_2)+e^{i\dlt}
 \bar{W}'(\bar{A}_{\rm cl}(x_2))}. 
\ee

The superfield equation of motion 
\be
 -\frac{1}{4}\bar{D}^2\bar{\Phi}+W'(\Phi)=0 
\ee
can be rewritten in terms of $\vph^{(1)}$ as \cite{sakamura1}
\be
 -\frac{1}{2}D_1^2\bar{\vph}^{(1)}+i\der_2\bar{\vph}^{(1)}
 -ie^{-i\dlt}W'(\vph^{(1)})=0. 
 \label{EOM_vph1}
\ee
By substituting $\vph^{(1)}=\vpcl^{(1)}+\tl{\vph}^{(1)}$ into 
Eq.(\ref{EOM_vph1}) and using the equation of motion for $\vpcl^{(1)}$, 
we can obtain the linearized equation of motion 
for the fluctuation field~$\tl{\vph}^{(1)}$ as 
\be
 -\frac{1}{2}D_1^2\bar{\tl{\vph}}^{(1)}+i\der_2\bar{\tl{\vph}}^{(1)}
 -ie^{-i\dlt}W''(\vpcl^{(1)})\tl{\vph}^{(1)}
 =0.  \label{lin_EOM_vph1}
\ee

Let us denote the component fields of $\tl{\vph}^{(1)}$ as 
\be
 \tl{\vph}^{(1)}=a^{(1)}+\tht_1\psi^{(1)}+\frac{1}{2}\tht_1^2f^{(1)}. 
\ee
Then, we can obtain the following linearized equation of motion 
for $\psi^{(1)}_\alpha$ by picking up linear terms for $\tht_1$ from 
Eq.(\ref{lin_EOM_vph1}). 
\be
 -i\brkt{\gm_{(3)}^m\der_m\bar{\psi}^{(1)}}_\alpha
 +i\brc{\der_2\bar{\psi}^{(1)}_\alpha-e^{-i\dlt}W''(\Acl)\psi^{(1)}_\alpha}
 =0. 
\ee
Therefore, the mode equation for $\psi^{(1)}_\alpha$ can be read off as 
\be
 i\brc{\der_2\bar{u}^{(1)}_{(n)}-e^{-i\dlt}W''(\Acl)u^{(1)}_{(n)}}
 =m_{(n)}\bar{u}^{(1)}_{(n)},  \label{mode_eq1}
\ee
where $m_{(n)}$ are eigenvalues. 
Note that this has a very similar form to that of the BPS-wall case 
\cite{sakamura1}. 
The only difference is that the argument of $W''$ is 
a non-BPS configuration~$\Acl$, instead of $\Acla$. 
On the other hand, the mode equation for $a^{(1)}$ receives more corrections 
than that for $\psi^{(1)}_\alpha$ when you change the background 
from the BPS one to the non-BPS one. 
Therefore, it is convenient to use the mode functions of 
$\psi^{(1)}_\alpha$ in the following derivation of the 3D effective theory. 

To simplify the discussion, we will consider a case that 
$\Acl(x_2)$ and all parameters of the 4D bulk theory are real. 
In this case, $\dlt=0$ \footnote{
There is another possibility that $\dlt=\pi$, but it can be reduced 
to the case of $\dlt=0$ by the redefinition of the coordinate: 
$x_2\to x_2+\pi R$. 
} and the mode equation~(\ref{mode_eq1}) can be 
rewritten as  
\bea
 \brc{-\der_2+W''(\Acl)}\uR{n} \eql m_{(n)}\uI{n}, \nonumber\\
 \brc{\der_2+W''(\Acl)}\uI{n} \eql m_{(n)}\uR{n},  \label{mode_eq2}
\eea
where $\uR{n}(x_2)$ and $\uI{n}(x_2)$ are defined by 
\be
 u^{(1)}_{(n)}=\frac{1}{\sqrt{2}}(\uR{n}+i\uI{n}). 
\ee
Using these mode functions, $\psi^{(1)}_\alpha$ can be expanded into 
3D fields as 
\be
 \psi^{(1)}_\alpha(x^m,x_2)=\frac{1}{\sqrt{2}}\sum_{n=0}^\infty
 \brkt{\uR{n}(x_2)\psi_{{\rm R}(n)\alpha}(x^m)
 +i\uI{n}(x_2)\psi_{{\rm I}(n)\alpha}(x^m)}. 
 \label{mode_expd_psi}
\ee

Here, note that the mode equation~(\ref{mode_eq2}) has two zero-modes, 
\bea
 \uR{0}(x_2) \eql C_0\,e^{\int_0^{x_2}\!{\rm d}y\:W''(\Acl(y))}, 
 \nonumber\\
 \uI{0}(x_2) \eql C_0\,e^{-\int_{\pi R}^{x_2}\!{\rm d}y\:W''(\Acl(y))}, 
 \label{zero_modes}
\eea
where $C_0$ is a common normalization factor. 
The corresponding fields~$\psR{0}$ and $\psI{0}$ are the goldstinos
for the broken $Q_2$- and $Q_1$-SUSY, and localized on the (approximate) 
BPS and anti-BPS walls, respectively. 
Therefore, it is convenient to introduce the following 3D superfields. 
\bea
 \vph^{(1)}_{(0)}(x^m,\tht_1) \eql a^{(1)}_{(0)}(x^m)+\tht_1\psR{0}(x^m)
 +\frac{1}{2}\tht_1^2f^{(1)}_{(0)}(x^m), \nonumber\\
 \vph^{(2)}_{(0)}(x^m,\tht_2) \eql a^{(2)}_{(0)}(x^m)+\tht_2\psI{0}(x^m)
 +\frac{1}{2}\tht_2^2f^{(2)}_{(0)}(x^m).  \label{ap_sf1}
\eea

Considering Eqs.(\ref{rel_vph12}) and (\ref{mode_expd_psi}), 
the mode-expansion of the fluctuation field~$\tl{\vph}^{(1)}$ 
can be expressed as 
\bea
 \tl{\vph}^{(1)}(x^m,x_2,\tht_1) \eql \frac{1}{\sqrt{2}}
 \brc{\uR{0}(x_2)\vph^{(1)}_{(0)}(x^m,\tht_1)+\uI{0}(x_2+i\tht_1^2)
 \vph^{(2)}_{(0)}(x^m,i\tht_1)} \nonumber\\
 &&+(\mbox{massive modes}).   \label{ap_mode-exp1}
\eea
Strictly speaking, the bosonic component of this expression deviates 
the correct mode-expansion 
because we used the mode functions of $\psi^{(1)}_\alpha$. 
However, this deviation is negligibly small when the distance 
between the walls is large enough\footnote{
In the BPS wall limit, the mode functions of bosonic and fermionic 
zero-modes approach to a common function.} 
On the other hand, the massive modes cannot be well described 
by using 3D superfields 
because they have broadly spread profiles along the $x_2$-direction 
and thus receive large SUSY breaking effects.

\subsection{Derivation of 3D effective theory}
In the following, we will concentrate ourselves on the low-energy region 
where all massive modes are decoupled, and will drop them from expressions. 
By substituting the expression 
\bea
 \vph^{(1)}(x^m,x_2,\tht_1) \eql \vpcl^{(1)}(x_2,\tht_1) \nonumber\\
 &&+\frac{1}{\sqrt{2}}\brc{\uR{0}(x_2)\vph^{(1)}_{(0)}(x^m,\tht_1)
 +\uI{0}(x_2+i\tht_1^2)\vph^{(2)}_{(0)}(x^m,i\tht_1)} 
\eea
into Eq.(\ref{S_vph1}) and carrying out the $x_2$-integration, 
we can obtain the 3D effective theory. 

A zero-th order term for the fields, {\it i.e.} the cosmological 
constant term, can be calculated as 
\bea
 \cL_{\rm cc}^{(3)} \eql \int\dr^2\tht_1{\rm d}x_2 \:
 \brc{D_1^\alpha\bar{\vph}_{\rm cl}^{(1)}D_{1\alpha}\vpcl^{(1)} 
 +\Im\brkt{2\bar{\vph}_{\rm cl}^{(1)}\der_2\vpcl^{(1)}
 +4W(\vpcl^{(1)})}} \nonumber\\
 \eql -\int\dr x_2\: \brc{(\der_2\Acl)^2+\brkt{W'(\Acl)}^2} =-V_0,
 \label{Lcc}
\eea
where $V_0$ is an energy (per unit area) of the background 
field configuration~$\Acl(x_2)$. 

Linear terms for the fields are cancelled due to 
the equation of motion for $\vpcl^{(1)}$. 

Then, we will calculate quadratic terms for the fields. 
\bea
 \cL_{\rm quad}^{(3)} \eql \int\dr^2\tht_1{\rm d}x_2\:
 \brc{D_1^\alpha\bar{\tl{\vph}}^{(1)}D_{1\alpha}\tl{\vph}^{(1)}
 -2\Im\brkt{\tl{\vph}^{(1)}\brkt{\der_2\bar{\tl{\vph}}^{(1)}
 -W''(\vpcl^{(1)})\tl{\vph}^{(1)}}}} \nonumber\\
 \eql \int\dr^2\tht_1\:\frac{1}{2}\brkt{D_1\vph^{(1)}_{(0)}}^2
 +\int\dr^2\tht_2\:\frac{1}{2}\brkt{D_2\vph^{(2)}_{(0)}}^2 \nonumber\\
 &&-\eta\brkt{f^{(1)}_{(0)}f^{(2)}_{(0)}
 +\der^ma^{(1)}_{(0)}\der_ma^{(2)}_{(0)}}
 -\frac{1}{2}m_0^2\brkt{a^{(1)}_{(0)}-a^{(2)}_{(0)}}^2, 
 \label{L_quad}
\eea
where $\eta \equiv \int{\rm d}x_2\: \uR{0}\uI{0}$ is 
a dimensionless number, which is exponentially suppressed 
in terms of the wall distance~$\pi R$, 
and the scalar mass parameter~$m_0^2$ is defined as 
\bea
 m_0^2 \defa -\int\dr x_2\: W'''(\Acl)\UI\uR{0}^2 
 =\int\dr x_2 \: W'''(\Acl)\UR\uI{0}^2 \nonumber\\
 \eql \int\dr x_2\: W'''(\Acl)\UI\uR{0}\uI{0} 
 =-\int\dr x_2\: W'''(\Acl)\UR\uR{0}\uI{0}, \label{def_m0}
\eea
Here, we have used the formulae Eqs.(\ref{K_vph1-vph2}) and 
(\ref{W_vph1-vph2}), 
and the functions~$\UR(x_2)$ and $\UI(x_2)$ are defined as 
\bea 
 \UR(x_2) \defa \der_2\Acl(x_2)+W'(\Acl(x_2)), \nonumber\\
 \UI(x_2) \defa \der_2\Acl(x_2)-W'(\Acl(x_2)),  \label{def_URI}
\eea
which are contained in $\vpcl^{(1)}$ and $\vpcl^{(2)}$ as 
\bea
 \vpcl^{(1)}(x_2,\tht_1)\eql \Acl(x_2)+\frac{i}{2}\tht_1^2\UI(x_2), 
 \nonumber\\
 \vpcl^{(2)}(x_2,\tht_2)\eql \Acl(x_2)+\frac{i}{2}\tht_2^2\UR(x_2). 
 \label{U_in_vpcl}
\eea
Note that $\UR(x_2)$ and $\UI(x_2)$ satisfy the mode equations~(\ref{mode_eq2}) 
with $m_{(n)}=0$, and are related to $\uR{0}(x_2)$ and $\uI{0}(x_2)$ as 
\be
 \UR(x_2)=\sqrt{V_0}\,\uR{0}(x_2), \;\;\;
 \UI(x_2)=\sqrt{V_0}\,\uI{0}(x_2). \label{rel_U-u}
\ee
Equalities in Eq.(\ref{def_m0}) are followed from 
this relation and Eqs.~(\ref{UR-UI_prdct}) and (\ref{WU_piRtrslt}) 
in Appendix~\ref{Acl_prop}. 

From Eqs.(\ref{Lcc}) and (\ref{L_quad}), 
the bosonic part of the 3D Lagrangian~$\cL^{(3)}$ is 
\bea
 \cL^{(3)}_{\rm bosonic} \eql -V_0-\frac{1}{2}
 (\der^ma^{(1)}_{(0)},\der^ma^{(2)}_{(0)})\mtrx{1}{\eta}{\eta}{1}
 \vct{\der_ma^{(1)}_{(0)}}{\der_ma^{(2)}_{(0)}} \nonumber\\
 &&-\frac{1}{2}(a^{(1)}_{(0)},a^{(2)}_{(0)})\mtrx{m_0^2}{-m_0^2}{-m_0^2}{m_0^2}
 \vct{a^{(1)}_{(0)}}{a^{(2)}_{(0)}} 
 +\frac{1}{2}(f^{(1)}_{(0)},f^{(2)}_{(0)})\mtrx{1}{-\eta}{-\eta}{1}
 \vct{f^{(1)}_{(0)}}{f^{(2)}_{(0)}} \nonumber\\
 &&+(\mbox{interaction terms}). 
 \label{Lboson}
\eea
Then, canonically normalized scalar fields are 
\bea
 a_{+(0)}\defa\sqrt{\frac{1+\eta}{2}}\brkt{a^{(1)}_{(0)}+a^{(2)}_{(0)}}, 
 \nonumber\\
 a_{-(0)}\defa\sqrt{\frac{1-\eta}{2}}\brkt{a^{(1)}_{(0)}-a^{(2)}_{(0)}}. 
 \label{def_apm}
\eea
For the auxiliary fields, we redefine them as 
\bea
 f_{+(0)}\defa\sqrt{\frac{1-\eta}{2}}\brkt{f^{(1)}_{(0)}+f^{(2)}_{(0)}}, 
\nonumber\\
 f_{-(0)}\defa\sqrt{\frac{1+\eta}{2}}\brkt{f^{(1)}_{(0)}-f^{(2)}_{(0)}}. 
\eea
Using these fields, Eq.(\ref{Lboson}) can be rewritten as 
\bea
 \cL^{(3)}_{\rm bosonic} \eql -V_0-\frac{1}{2}\der^ma_{+(0)}\der_ma_{+(0)}
 -\frac{1}{2}\der^ma_{-(0)}\der_ma_{-(0)}
 -\frac{1}{2}\brkt{\frac{2m_0^2}{1-\eta}}a_{-(0)}^2 \nonumber\\
 &&+\frac{1}{2}f_{+(0)}^2+\frac{1}{2}f_{-(0)}^2+(\mbox{interaction terms}). 
 \label{Lboson2}
\eea

Notice that the massless mode $a_{+(0)}(x^m)$ is the Nambu-Goldstone mode 
for the translation along the $x_2$-direction. 
In fact, the mode function of $a_{+(0)}(x^m)$ is 
\bea
 u_{+(0)}(x_2) \eql \frac{1}{\sqrt{2(1-\eta)}}\brkt{\uR{0}(x_2)+\uI{0}(x_2)}
 \nonumber\\
 &&\frac{1}{\sqrt{2(1-\eta)V_0}}\brkt{\UR(x_2)+\UI(x_2)}
 =\sqrt{\frac{2}{(1-\eta)V_0}}\,\der_2\Acl(x_2). 
\eea
This mode function is correctly normalized as expected. 
\be
 \int\dr x_2\: u_{+(0)}^2 =1. 
\ee

We can calculate interaction terms in a similar way. 
For example, cubic couplings are obtained as 
\bea 
 \cL^{(3)}_{\rm cubic} \eql \int\dr^2\tht_1{\rm d}x_2\:
 4\Im\brc{\frac{1}{3!}W'''(\vpcl^{(1)})\tl{\vph}^3} \nonumber\\
 \eql \cA_{111}\brkt{a^{(1)}_{(0)}}^3+\cA_{112}\brkt{a^{(1)}_{(0)}}^2a^{(2)}_{(0)}
 +\cA_{122}a^{(1)}_{(0)}\brkt{a^{(2)}_{(0)}}^2+\cA_{222}\brkt{a^{(2)}_{(0)}}^3 
 \nonumber\\
 &&+\frac{g_{(-{\rm RI})}}{\sqrt{2}}\brkt{a^{(1)}_{(0)}-a^{(2)}_{(0)}}\psR{0}\psI{0}, 
 \label{L_cubic}
\eea
where 
\bea
 \cA_{111} \defa \frac{1}{6\sqrt{2}}\int\dr x_2\: W'''(\Acl)\UI\uR{0}^3, 
 \nonumber\\
 \cA_{112} \defa \frac{1}{\sqrt{2}}\int\dr x_2\: 
 \brc{W'''(\Acl)\uR{0}^2\der_2\uI{0}+\frac{1}{2}W''''(\Acl)\UI\uR{0}^2\uI{0}}, 
 \nonumber\\
 \cA_{122} \defa \frac{1}{\sqrt{2}}\int\dr x_2\:
 \brc{W'''(\Acl)\uI{0}^2\der_2\uR{0}+\frac{1}{2}W''''(\Acl)\UR\uI{0}^2\uR{0}}, 
 \nonumber\\
 \cA_{222} \defa \frac{1}{6\sqrt{2}}\int\dr x_2\: W'''(\Acl)\UR\uI{0}^3, 
 \label{def_cA} \\
 g_{(-{\rm RI})} \defa -\int\dr x_2\: W'''(\Acl)\uR{0}^2\uI{0}
 =\int\dr x_2\: W'''(\Acl)\uI{0}^2\uR{0}.  \label{def_gef}
\eea
The second equality in Eq.(\ref{def_gef}) is followed from 
the relations~(\ref{rel_U-u}), (\ref{UR-UI_prdct}) and (\ref{WU_piRtrslt}). 

The ``$A$-terms'' for $\cA_{111}$ and $\cA_{222}$ correspond to Eq.(3.3) 
in Ref.\cite{MSSS}, but terms for $\cA_{112}$ and $\cA_{122}$ 
were missed there. 

Notice that the above cubic couplings are all SUSY breaking terms. 
In fact, we can easily show that there are no supersymmetric couplings 
in the effective theory. 
This stems from the assumption that $\Acl(x_2)$ and all parameters 
in the bulk theory are real. 
As will be seen in the next section, supersymmetric interaction terms 
appear in the effective theory if we introduce a complex coupling constant 
in the bulk theory. 

The existence of the last Yukawa coupling term in Eq.(\ref{L_cubic}) 
is expected from the low-energy theorem \cite{lee,clark}, which predicts 
the relation among the Yukawa coupling constant~$g_{(-RI)}$,  
the mass-splitting between the bosonic and fermionic modes~$m_0^2$ 
and an order parameter of SUSY breaking~$\sqrt{V_0}$ as 
\be
 g_{(-{\rm RI})}=\frac{m_0^2}{\sqrt{V_0}}. 
\ee
This relation can easily be seen from Eqs.(\ref{def_m0}), (\ref{rel_U-u}) 
and (\ref{def_gef}) in our approach.

\subsection{Matter fields} \label{matter}
In order to construct a more realistic model, we need to introduce 
matter fields, which do not contribute to the domain wall configuration. 
Here, we will introduce an extra chiral superfield~$X$ as matter fields 
whose interactions are given by 
\be
 W_{\rm int}(\Phi,X)=\frac{\alpha}{2}W''(\Phi)X^2+\beta X^3, 
 \label{Wm_int}
\ee
where constants $\alpha$ and $\beta$ are positive and complex, respectively. 
The first term in Eq.(\ref{Wm_int}) is required in order to localize 
the lowest modes of $X$ on the walls, and the second term provides 
interactions among the matter fields, which are important 
when we discuss the phenomenology\footnote{
Proliferation of the matter fields is straightforward. 
}. 
The component fields of $X$ are denoted as 
\be
 X(y,\tht)=A_X(y)+\sqrt{2}\tht\Psi_X(y)+\tht^2F_X(y). 
\ee
We will assume that $A_X(x)$ does not have a nontrivial background, 
{\it i.e.} $A_{X{\rm cl}}=0$.
The corresponding field to $\vph^{(1)}$ in Eq.(\ref{def_vph1}) is 
\bea
 \vph^{(1)}_X(x^m,x_2,\tht_1) \defa e^{\tht_1Q_1}
 \times A_X(x^m,x_2)  \nonumber\\
 \eql a^{(1)}_X(x^m,x_2)+\tht_1\psi^{(1)}_X(x^m,x_2)
 +\frac{1}{2}\tht_1^2f^{(1)}_X(x^m,x_2).
\eea

The mode equations for the fermionic component~$\psi^{(1)}_{X\alpha}$ 
can be obtained in the same way as the derivation of Eq.(\ref{mode_eq2}).  
\bea
 \brc{-\der_2+\alpha W''(\Acl)}\uRX{n} \eql m_{X(n)}\uIX{n}, \nonumber\\
 \brc{\der_2+\alpha W''(\Acl)}\uIX{n} \eql m_{X(n)}\uRX{n}, 
 \label{mode_eq3}
\eea
where $m_{X(n)}$ are eigenvalues. 
Using solutions of these equations, $\psi^{(1)}_{X\alpha}$ can be expanded as 
\be
 \psi^{(1)}_{X\alpha}(x^m,x_2)=\frac{1}{\sqrt{2}}\sum_{n=0}^\infty 
 \brc{\uRX{n}(x_2)\psi_{{\rm R}(n)\alpha}(x^m)
 +i\uIX{n}(x_2)\psi_{{\rm I}(n)\alpha}(x^m)}. 
\ee

From Eq.(\ref{mode_eq3}), we can see that there exist two zero-modes 
whose mode functions are  
\bea
 \uRX{0}(x_2) \eql C_Xe^{\alpha\int_0^{x_2}\!{\rm d}y\:
 W''(\Acl(y))},  \nonumber\\
 \uIX{0}(x_2) \eql C_Xe^{-\alpha\int_{\pi R}^{x_2}\!{\rm d}y\:
 W''(\Acl(y))},  \label{zero_modes2}
\eea
where $C_X$ is a common normalization factor. 

Following Eqs.(\ref{ap_sf1}) and (\ref{ap_mode-exp1}), 
we can see that $\vph^{(1)}_X$ can be expanded approximately as
\bea
 \vph^{(1)}_X \eql \frac{1}{\sqrt{2}}
 \brc{\uRX{0}(x_2)\vph^{(1)}_{X(0)}(x^m,\tht_1)
 +\uIX{0}(x_2+i\tht_1^2)\vph^{(2)}_{X(0)}(x^m,i\tht_1)} \nonumber\\
 && +(\mbox{massive modes}),  \label{ap_mode-exp2}
\eea
where 
\bea
 \vph^{(1)}_{X(0)}(x^m,\tht_1) \eql a^{(1)}_{X(0)}(x^m)
 +\tht_1\psRX{0}(x^m)+\frac{1}{2}\tht_1^2f^{(1)}_{X(0)}(x^m), 
 \nonumber\\
 \vph^{(2)}_{X(0)}(x^m,\tht_2) \eql a^{(2)}_{X(0)}(x^m)
 +\tht_2\psIX{0}(x^m)+\frac{1}{2}\tht_2^2f^{(2)}_{X(0)}(x^m).
\eea

Then, after the massive modes are decoupled, we can obtain 
the effective Lagrangian for the matter fields by substituting 
Eq.(\ref{ap_mode-exp2}) into the original 4D Lagrangian and 
carrying out the $x_2$-integration. 

The quadratic part for the fields is 
\bea
 \cL^{(3)X}_{\rm quad} \eql \int{\rm d}^2\tht_1 \:
 \frac{1}{2}\brkt{D_1\vph^{(1)}_{X(0)}}^2
 +\int{\rm d}^2\tht_2 \:
 \frac{1}{2}\brkt{D_2\vph^{(2)}_{X(0)}}^2
 -\eta_X\brkt{f^{(1)}_{X(0)}f^{(2)}_{X(0)}
 +\der^ma^{(1)}_{X(0)}\der_ma^{(2)}_{X(0)}} \nonumber\\
 && -\frac{1}{2}(a_{X(0)}^{(1)},a_{X(0)}^{(2)})
 \mtrx{m_X^2}{-\Dlt m_X^2}{-\Dlt m_X^2}{m_X^2}
 \vct{a_{X(0)}^{(1)}}{a_{X(0)}^{(2)}}, 
 \label{L_matter}
\eea
where 
\bea
 \eta_X \defa \int\dr x_2\: \uRX{0}\uIX{0}, \nonumber\\
 m_X^2 \defa -\alpha\int\dr x_2 \: W'''(\Acl)\UI\uRX{0}^2 
 =\alpha\int\dr x_2\: W'''(\Acl)\UR\uIX{0}^2,  \\
 \Dlt m_X^2 \defa \alpha\int\dr x_2 \: W'''(\Acl)\UI\uRX{0}\uIX{0} 
 =-\alpha\int\dr x_2 \: W'''(\Acl)\UR\uRX{0}\uIX{0}.  \nonumber
\eea
From Eq.(\ref{L_matter}), the mass eigenstates of the scalar fields are 
\bea
 a_{X+(0)}\defa \sqrt{\frac{1+\eta_X}{2}}\brkt{a^{(1)}_{X(0)}+a^{(2)}_{X(0)}}, 
 \nonumber\\
 a_{X-(0)}\defa \sqrt{\frac{1-\eta_X}{2}}\brkt{a^{(1)}_{X(0)}-a^{(2)}_{X(0)}}, 
\eea
whose mass eigenvalues are 
\bea
 m_{X+}^2 \eql \frac{m_X^2-\Dlt m_X^2}{1+\eta_X}, \nonumber\\
 m_{X-}^2 \eql \frac{m_X^2+\Dlt m_X^2}{1-\eta_X}.  
\eea

For different values of $\alpha$, the mass spectrum of the scalar modes 
has different features. 
First, let us consider the case of $\alpha\gg 1$.\footnote{
$\alpha\gg 1$ does not necessarily mean the strong coupling region. 
For example, see the comment below Eq.(\ref{DmX}).}
In this case, $\uRX{0}(x_2)$ has a sharp peak around $x_2=0$ 
and can be well approximated there by the following Gaussian function. 
\be
 \uRX{0}(x_2) \simeq C_Xe^{-\frac{\alpha b}{2}x_2^2}, \;\;\;
 (\mbox{in the neighborhood of $x_2=0$.})
\ee
where $b\equiv -W'''(\Acl(0))\der_2\Acl(0)$.\footnote{
From the assumption that $\dlt=0$, we can show that $b>0$. }
This means that the normalization factor~$C_X$ is estimated as 
\be
 C_X \simeq \brkt{\frac{\alpha b}{\pi}}^{1/4}. 
\ee
Therefore, we can approximate $\uRX{0}(x_2)$ as 
\be
 \uRX{0}(x_2) \simeq \brkt{\frac{\alpha b}{\pi}}^{1/4} 
 \brkt{\frac{\UR(x_2)}{\UR(0)}}^\alpha. 
\ee
As a result, we can clarify an $\alpha$-dependence of $\Dlt m_X^2$ 
as follows. 
\be
 \Dlt m_X^2 \simeq \sqrt{\frac{\alpha^3 b}{\pi}}
 \brkt{\frac{V_0}{\UR(0)^2}\frac{\eta}{2\pi R}}^\alpha 
 \int\dr x_2\: W'''(\Acl)\UI.  \label{alph_dep_Dms}
\ee
Here we have used the formulae~(\ref{UR-UI_prdct}) and (\ref{c_expl}) 
in Appendix~\ref{Acl_prop}. 
From this expression, we can see that the absolute value of $\Dlt m_X^2$ 
decreases exponentially as $\alpha$ increases. 

On the other hand, $m_X^2$ increases linearly as $\alpha$ grows up\footnote{
We can show that R.H.S. in Eq.(\ref{limit_mXs}) is positive 
from the assumption that $\dlt=0$. 
}. 
\be
 m_X^2 \simeq -\alpha W'''(\Acl(0))\UI(0), \;\;\;
 (\mbox{when $\alpha\gg 1$}) \label{limit_mXs}
\ee
because $\uRX{0}^2(x_2)$ can be approximated by $\dlt(x_2)$ 
when $\alpha$ is large enough. 

From Eqs.(\ref{alph_dep_Dms}) and (\ref{limit_mXs}), 
we can conclude that the two lightest modes~$a_{X\pm(0)}$ 
become almost degenerate with a large mass eigenvalue 
compared to $m_0^2$ in Eq.(\ref{def_m0}) 
in the case of $\alpha\gg 1$. 

In the case of $\alpha=1$, the mode equation~(\ref{mode_eq3}) has 
the same form as Eq.(\ref{mode_eq2}). 
As a result, $m_X^2=\Dlt m_X^2$ and a massless scalar besides 
the NG mode~$a_{+(0)}$ appears in the 3D effective theory. 

As we will see in Sect.\ref{exmpl_model}, in the case of $0<\alpha < 1$, 
it follows that $m_X^2<|\Dlt m_X^2|$ and thus 
there appears a tachyonic mode in the effective theory. 
This means that the classical background~$(A,A_X)=(\Acl(x_2),0)$ is 
unstable. 

In the case of $\alpha=0$, both $m_X^2$ and $\Dlt m_X^2$ become zero, 
and thus both scalar modes~$a_{X\pm(0)}$ are massless. 
This is a trivial result because 
SUSY breaking in the $\Phi$-sector is not transmitted 
to the $X$-sector at tree-level in this case. 

Next, we will consider interaction terms. 
They are calculated just in a similar way to the derivation 
of Eq.(\ref{L_cubic}). 
The interaction terms among the matter fields, which come from 
the second term of Eq.(\ref{Wm_int}), are as follows. 
\bea
 \cL^{(3){\rm m}}_{\rm cubic} \eql \int\dr^2\tht_1\: 
 \befa\brkt{\vph_{X(0)}^{(1)}}^3
 +\int\dr^2\tht_2\: \befb\brkt{\vph_{X(0)}^{(2)}}^3 \nonumber\\
 &&+(\ep_{X1}\Im\beta)\brc{2a_{X(0)}^{(1)}a_{X(0)}^{(2)}f_{X(0)}^{(1)}
 -\brkt{a_{X(0)}^{(1)}}^2f_{X(0)}^{(2)}-a_{X(0)}^{(2)}\psRX{0}^2} 
 \nonumber\\
 &&-(\ep_{X1}\Re\beta)a_{X(0)}^{(1)}\psRX{0}\psIX{0} \nonumber\\
 &&+(\ep_{X2}\Im\beta)\brc{2a_{X(0)}^{(1)}a_{X(0)}^{(2)}f_{X(0)}^{(2)} 
 -\brkt{a_{X(0)}^{(2)}}^2f_{X(0)}^{(1)}-a_{X(0)}^{(1)}\psIX{0}^2} 
 \nonumber\\
 &&-(\ep_{X2}\Re\beta)a_{X(0)}^{(2)}\psRX{0}\psIX{0} \nonumber\\
 &&+\cA_{112}^X \brkt{a_{X(0)}^{(1)}}^2a_{X(0)}^{(2)}
 +\cA_{122}^X \cdot a_{X(0)}^{(1)}\brkt{a_{X(0)}^{(2)}}^2,
 \label{L_cubic_mat}
\eea
where 
\bea
 \befa\defa \frac{\Im\beta}{3\sqrt{2}}\int\dr x_2\: \uRX{0}^3, \;\;\;
 \befb\equiv \frac{\Im\beta}{3\sqrt{2}}\int\dr x_2\: \uIX{0}^3, \nonumber\\
 \ep_{X1}\defa \frac{1}{\sqrt{2}}\int\dr x_2\: \uRX{0}^2\uIX{0}, \;\;\;
 \ep_{X2}\equiv \frac{1}{\sqrt{2}}\int\dr x_2\: \uRX{0}\uIX{0}^2, 
 \nonumber\\
 \cA_{112}^X \defa \frac{\Re\beta}{\sqrt{2}}\int\dr x_2\: 
 \uRX{0}^2\der_2\uIX{0},  \;\;\;
 \cA_{122}^X \equiv \frac{\Re\beta}{\sqrt{2}}\int\dr x_2\:
 \der_2\uRX{0}\uIX{0}^2. 
\eea
The first line of R.H.S. in Eq.(\ref{L_cubic_mat}) represents 
supersymmetric interactions, which were absent in Eq.(\ref{L_cubic}), 
and the remaining terms are 
SUSY breaking terms.  

The interaction terms between the ``wall fields'' and the matter fields, 
which come from the first term of Eq.(\ref{Wm_int}), can also 
be calculated in a similar way.

\section{BPS-wall approximation} \label{BPSapx}
In this section, we will provide a useful approximation 
for the derivation of the 3D effective theory\footnote{
A method we will provide in this section is essentially the same one 
as the ``single-wall approximation'' proposed in Ref.\cite{MSSS},  
but it was used only to estimate the mass-splittings 
between bosons and fermions. 
Here, we will show that it can be applied to the estimation 
of {\it all parameters} in the effective theory.   
}. 
Since the BPS equations~(\ref{BPS_eq1}) and (\ref{BPS_eq2}) are 
first order differential equations, 
finding the BPS wall solutions~$\Acla(x_2)$ and $\Aclb(x_2)$ is 
easier than finding the non-BPS configuration~$\Acl(x_2)$, 
which is a solution of a second order differential equation. 
So it is a practical method 
to approximate $\Acl(x_2)$ by using $\Acla(x_2)$ and $\Aclb(x_2)$. 

For example, let us derive an approximate expression of $\UI(x_2)$ 
defined in Eq.(\ref{def_URI}). 

Notice that $\Acl(x_2)$ is well approximated as follows\footnote{
Locations of the walls are taken to be at $x_2=0$ in the definition 
of $\Acla(x_2)$ and $\Aclb(x_2)$. 
}. 
\be
 \Acl(x_2)\simeq\left\{\begin{array}{l}\Acla(x_2) \;\;\; 
 \brkt{-\frac{\pi R}{2}<x_2\leq\frac{\pi R}{2}} \\
 \Aclb(x_2-\pi R) \;\;\; 
 \brkt{\frac{\pi R}{2}<x_2\leq\frac{3\pi R}{2}}\end{array}\right. 
\ee
Then, using the BPS equation~(\ref{BPS_eq2}), 
\be
 \UI(x_2)\simeq 2\der_2\Aclb(x_2-\pi R), \label{UIap1}
\ee
in the region of $\pi R/2<x_2\leq 3\pi R/2$. 

On the other hand, from Eqs.(\ref{zero_modes}) and (\ref{rel_U-u}), 
$\UI(x_2)$ can be written as 
\be
 \UI(x_2)=\UI(\pi R)e^{-\int_{\pi R}^{x_2}{\rm d}y\: W''(\Acl(y))}. 
 \label{UIexp}
\ee
From Eq.(\ref{UIap1}), the constant factor~$\UI(\pi R)$ is estimated as 
\be
 \UI(\pi R)\simeq 2\der_2\Aclb(0). \label{UI_piR}
\ee
Thus, in the region of $-\pi R/2<x_2\leq \pi R/2$, Eq.(\ref{UIexp}) is 
approximated as 
\bea
 \UI(x_2)\eql \UI(\pi R)e^{-\int_0^{x_2}{\rm d}y\: W''(\Acl(y))
 +\int_0^{\pi R/2}{\rm d}y\: W''(\Acl(y))+\int_{\pi R/2}^{\pi R}{\rm d}y\: 
 W''(\Acl(y))} \nonumber\\
 \eql \UI(\pi R)e^{2\int_0^{\pi R/2}{\rm d}y\: W''(\Acl(y))}
 e^{-\int_0^{x_2}{\rm d}y\: W''(\Acl(y))} \nonumber\\
 \!\!\!&\simeq \!\!\!& 2\der_2\Aclb(0)e^{2\int_0^{\pi R/2}{\rm d}y\: 
 W''(\Acla(y))}\cdot e^{-\int_0^{x_2}{\rm d}y\: W''(\Acla(y))}. 
\eea
Here we have used Eq.(\ref{Wpp_prop}) in Appendix~\ref{Acl_prop}. 

Consequently, we can approximate $\UI(x_2)$ as 
\be
 \UI(x_2)\simeq \left\{\begin{array}{l}
 2\der_2\Aclb(0)e^{2\int_0^{\pi R/2}{\rm d}y\: W''(\Acla(y))}\cdot
 e^{-\int_0^{x_2}{\rm d}y\: W''(\Acla(y))}, \;\;\;
 \brkt{-\frac{\pi R}{2}<x_2\leq\frac{\pi R}{2}} \\
 2\der_2\Aclb(x_2-\pi R). \;\;\; 
 \brkt{\frac{\pi R}{2}<x_2\leq\frac{3\pi R}{2}} \end{array}\right. 
 \label{apx_UI}
\ee
An approximate expression of $\UR(x_2)$ can be obtained in a similar way. 
\be
 \UR(x_2)\simeq \left\{\begin{array}{l}
 2\der_2\Acla(x_2). \;\;\; 
 \brkt{-\frac{\pi R}{2}<x_2\leq\frac{\pi R}{2}} \\
 2\der_2\Acla(0)e^{2\int_0^{\pi R/2}{\rm d}y\: W''(\Acla(y))}\cdot
 e^{\int_{\pi R}^{x_2}{\rm d}y\: W''(\Aclb(y-\pi R))}, \;\;\;
 \brkt{\frac{\pi R}{2}<x_2\leq\frac{3\pi R}{2}} 
 \end{array}\right. 
\ee
Note that these expressions are continuous at $x_2=\pm\pi R/2$. 

The energy (per unit area) of the configuration~$V_0$ can be well approximated 
by the sum of the tension of the two BPS walls. 
\bea
 V_0 \!\!\!&\simeq\!\!\!& \int\dr x_2\: \brc{(\der_2\Acla)^2+(W'(\Acla))^2}
 +\int\dr x_2\: \brc{(\der_2\Aclb)^2+(W'(\Aclb))^2} \nonumber\\
 \eql 2\int\dr x_2\: \brc{(\der_2\Acla)^2+(W'(\Acla))^2}. 
\eea

From these expressions, we can also obtain approximate expressions of 
the mode functions~$\uR{0}(x_2)$, $\uI{0}(x_2)$, $\uRX{0}(x_2)$ 
and $\uIX{0}(x_2)$ through Eqs.(\ref{rel_U-u}) and (\ref{zero_modes2}). 
Hence, using all of them, we can calculate all parameters of 
the 3D effective theory with high accuracy.

\section{Example of wall-antiwall configuration} \label{exmpl_model}
\begin{figure}[t]
 \leavevmode
 \epsfysize=7cm
 \centerline{\epsfbox{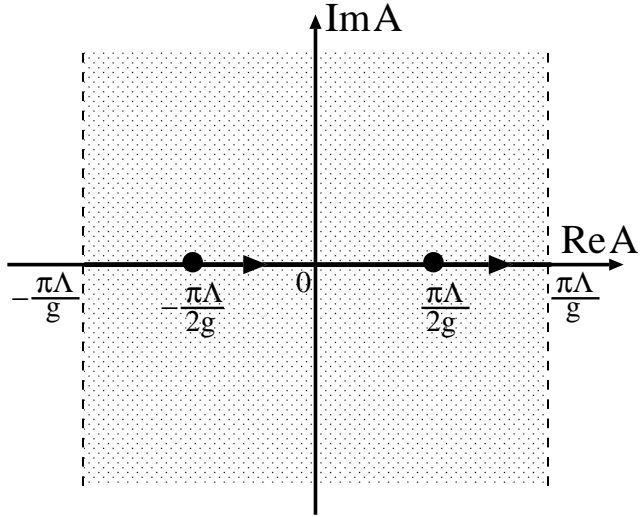}}
 \caption{The target space of the scalar field $A$. The lines at 
 $\Re A=\pm\pi\Lmd/g$ are identified with each other. The SUSY vacua 
 are $A=\pm\pi\Lmd/(2g)$. }
 \label{target-A}
\end{figure}

In this section, we will provide an example of the wall-antiwall 
configuration in a specific model, and demonstrate explicit calculations 
discussed in the previous sections. 
The model we consider here is the following sine-Gordon type model 
\cite{MSSS}. 
\be
 \cL=\bar{\Phi}\Phi|_{\tht^2\btht^2}+W(\Phi)|_{\tht^2}+{\rm h.c.}, \;\;\;
 W(\Phi)=\frac{\Lmd^3}{g^2}\sin\brkt{\frac{g}{\Lmd}\Phi}, 
\ee
where a scale parameter~$\Lmd$ has a mass-dimension one and 
a coupling constant~$g$ is dimensionless, and both of them 
are real positive. 
The bosonic part of the model is 
\be
 \cL_{\rm bosonic}=-\der^\mu\bar{A}\der_\mu A-\frac{\Lmd^4}{g^2}
 \left |\cos\brkt{\frac{g}{\Lmd}A}\right |^2 .
\ee
The target space of the scalar field~$A$ has a topology of 
a cylinder as shown in Fig.\ref{target-A}. 
This model has two supersymmetric vacua at $A=\pm\frac{\pi\Lmd}{2g}$, 
both lie on the real axis. 

There are BPS and anti-BPS walls interpolating these two vacua. 
The BPS wall solution is 
\be
 \Acla(x_2)=\frac{\Lmd}{g}\brc{2\tan^{-1}e^{\Lmd x_2}-\frac{\pi}{2}}, 
\ee
which interpolates from the vacuum at $\AI=-\frac{\pi\Lmd}{2g}$ to 
the one at $\AII=\frac{\pi\Lmd}{2g}$ 
as $y$ increases from $y=-\infty$ to $y=\infty$. 
The anti-BPS wall solution is 
\be
 \Aclb(x_2)=\frac{\Lmd}{g}\brc{-2\tan^{-1}e^{-\Lmd x_2}+\frac{3\pi}{2}}, 
\ee
which interpolates from $\AII=\frac{\pi\Lmd}{2g}$ to 
$\AI=\frac{3\pi\Lmd}{2g}(=-\frac{\pi\Lmd}{2g})$. 
Thus, a field configuration corresponding to the coexistence of these walls 
will wrap around the cylinder of the target space of $A$ 
as $x_2$ goes around $S^1$. 
Such a topological winding ensures the stability of the configuration. 

\begin{figure}[t]
 \leavevmode
 \epsfysize=8cm
 \centerline{\epsfbox{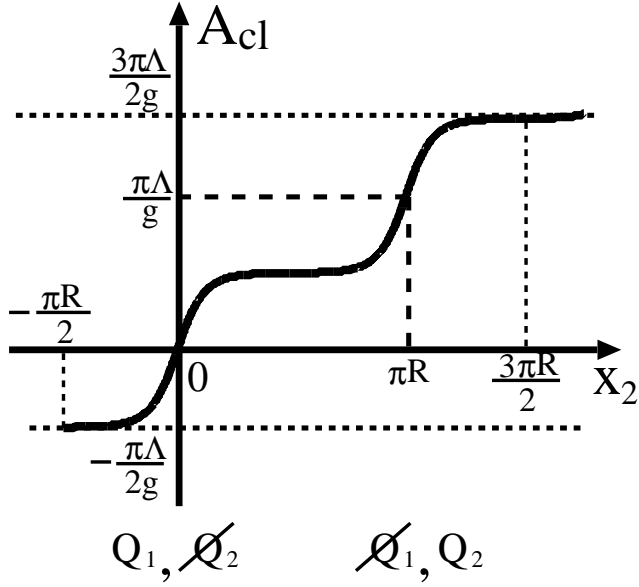}}
 \caption{The profile of the classical solution $\Acl(x_2)$. 
 The dotted lines $A=-\pi\Lmd/(2g)$ and $A=3\pi\Lmd/(2g)$ 
 are identified.
 }
 \label{profile-Acl}
\end{figure}

Indeed, there is a solution of the equation of motion, 
\be
 \der^\mu\der_\mu A+\frac{\Lmd^3}{g}\sin\brkt{\frac{g}{\Lmd}\bar{A}}
 \cos\brkt{\frac{g}{\Lmd}A}=0,  \label{EOM}
\ee
which corresponds to the wall-antiwall system. 
That is, 
\be
 \Acl(x_2)=\frac{\Lmd}{g}{\rm am}\brkt{\frac{\Lmd}{k}x_2,k}, 
 \label{exact_sol}
\ee
where $k$ is a real parameter and the function ${\rm am}(u,k)$ denotes 
the amplitude function, which is defined as an inverse function of 
\be
 u(\vph)=\int_0^\vph\: \frac{{\rm d}\tht}{\sqrt{1-k^2\sin^2\tht}}. 
\ee
For $k<1$, $\Acl(x_2)$ is a monotonically increasing function, 
and has a desired topological winding number. 
The parameter~$k$ is related to the radius of the compact space~$R$ 
through 
\be
 2\pi R=\frac{4k}{\Lmd}K(k), 
\ee
where $K(k)$ is the complete elliptic integral of the first kind. 
In the case of $2\pi R\Lmd\gg 1$, in other words, 
when the configuration~$\Acl(x_2)$ can be regarded as a wall-antiwall system, 
the parameter~$k$ is close to one, and can approximately be written as 
\be
 k\simeq \sqrt{1-16e^{-\pi R\Lmd}}. 
\ee
A profile of $\Acl(x_2)$ is shown in Fig.\ref{profile-Acl}.

\begin{figure}[t]
 \leavevmode
 \epsfysize=6cm
 \centerline{\epsfbox{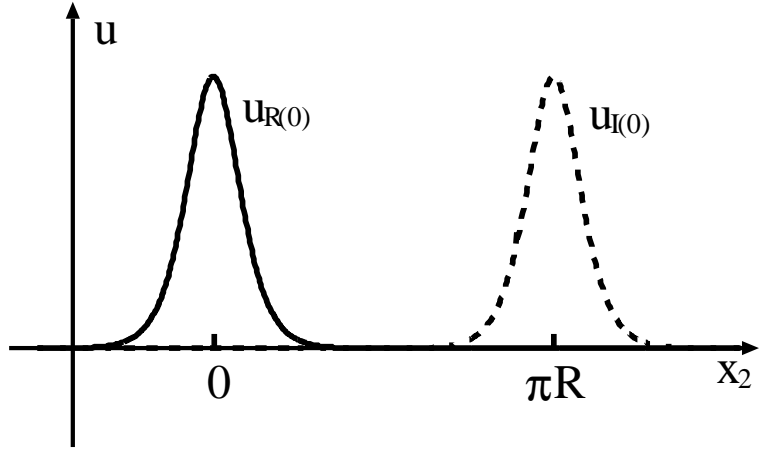}}
 \caption{The profiles of the mode-functions~$\uR{0}(x_2)$ 
  and $\uI{0}(x_2)$.}
 \label{mode_func}
\end{figure}

Solving the mode equation~(\ref{mode_eq2}) for this background, 
we can obtain the normalized mode functions of the zero-modes as follows. 
\bea
 \uR{0}(x_2) \eql \frac{\Lmd^2}{gk\sqrt{V_0}}
 \brc{\dn\brkt{\frac{\Lmd}{k}x_2,k}+k\cn\brkt{\frac{\Lmd}{k}x_2,k}}, 
 \nonumber\\
 \uI{0}(x_2) \eql \frac{\Lmd^2}{gk\sqrt{V_0}}
 \brc{\dn\brkt{\frac{\Lmd}{k}x_2,k}-k\cn\brkt{\frac{\Lmd}{k}x_2,k}}, 
\eea
where functions~$\cn(u,k)$ and $\dn(u,k)$ are the Jacobi's elliptic 
functions. 
The profiles of these mode-functions are shown in Fig.\ref{mode_func}. 
With these expressions, we can calculate all parameters of 
the 3D effective theory by using the formulae~(\ref{def_m0}), 
(\ref{def_cA}) and (\ref{def_gef}). 

As discussed in Sect.\ref{matter}, we can introduce matter fields localized 
on the walls. 
Let us introduce a matter chiral superfield~$X$ in the 4D bulk theory, 
and assume an interaction to the ``wall superfield''~$\Phi$ as 
\be 
 W_{\rm int}(\Phi,X)=\frac{\alpha}{2}W''(\Phi)X^2
 =-\frac{\alpha}{2}\Lmd\sin\brkt{\frac{g}{\Lmd}\Phi}X^2, 
 \label{W_mint_exp}
\ee
where a dimensionless constant~$\alpha$ is positive. 

The field configuration
\bea
 \Acl(x_2)\eql \frac{\Lmd}{g}{\rm am}\brkt{\frac{\Lmd}{k}x_2,k}, 
 \nonumber\\
 A_{X{\rm cl}}(x_2) \eql 0,  \label{bkgd}
\eea
is a static solution of the equation of motion. 
Solving the mode equations~(\ref{mode_eq3}) for this background, 
we can obtain the mode functions of the zero-modes for the matter fields 
as follows. 
\bea
 \uRX{0}(x_2)\eql C_X \brc{\dn\brkt{\frac{\Lmd}{k}x_2,k}
 +k\cn\brkt{\frac{\Lmd}{k}x_2,k}}^\alpha, \nonumber\\
 \uIX{0}(x_2))\eql C_X \brc{\dn\brkt{\frac{\Lmd}{k}x_2,k}
 -k\cn\brkt{\frac{\Lmd}{k}x_2,k}}^\alpha, 
\eea
where $C_X$ is a common normalization factor\footnote{
The definition of $C_X$ here is different from that in Eq.(\ref{zero_modes2}) 
by a factor of $(1+k)^\alpha$. 
}. 
Then, the 3D effective theory in this case is 
\bea
 \cL^{(3)}\eql -V_0+\int\dr^2\tht_1\:
 \brc{\frac{1}{2}(D_1\vph^{(1)}_{(0)})^2
 +\frac{1}{2}(D_1\vph^{(1)}_{X(0)})^2}
 +\int\dr^2\tht_2\:
 \brc{\frac{1}{2}(D_2\vph^{(2)}_{(0)})^2
 +\frac{1}{2}(D_2\vph^{(2)}_{X(0)})^2}
 \nonumber\\
 &&-\eta\brkt{f^{(1)}_{(0)}f^{(2)}_{(0)}
 +\der^ma^{(1)}_{(0)}\der_ma^{(2)}_{(0)}}
 -\eta_X\brkt{f^{(1)}_{X(0)}f^{(2)}_{X(0)}
 +\der^ma^{(1)}_{X(0)}\der_ma^{(2)}_{X(0)}} 
\nonumber\\
 &&-\frac{1}{2}m_0^2\brkt{a^{(1)}_{(0)}-a^{(2)}_{(0)}}^2 
 -\frac{1}{2}m_X^2\brkt{(a^{(1)}_{X(0)})^2+(a^{(2)}_{X(0)})^2}
 -\Dlt m_X^2 a^{(1)}_{X(0)}a^{(2)}_{X(0)} \nonumber\\
 &&+\mbox{(interaction terms)},
\eea
where 
\bea
 m_0^2 \eql \frac{1-k^2}{2k^2}\Lmd^2\brkt{1-\frac{2(1-k^2)}{g^2k^2V_0}
 \Lmd^4\pi R}, \nonumber\\
 m_X^2 \eql \alpha\frac{\Lmd^2}{k}C_X^2(1-k^2)\int\dr x_2 \:
 \cn\brkt{\frac{\Lmd}{k}x_2,k}\brc{\dn\brkt{\frac{\Lmd}{k}x_2,k}
 +k\cn\brkt{\frac{\Lmd}{k}x_2,k}}^{2\alpha-1}, \nonumber\\
 \Dlt m_X^2 \eql \alpha\Lmd^2C_X^2(1-k^2)^\alpha 
 \brkt{\frac{g^2V_0}{2\Lmd^4}-\frac{1-k^2}{k^2}\pi R}, \nonumber\\
 \eta \eql \frac{2\pi R}{V_0}\frac{\Lmd^4}{g^2k^2}(1-k^2), \;\;\;\;\;\;\;
 \eta_X = 2\pi R C_X^2(1-k^2)^\alpha. 
 \label{DmX}
\eea
From these expressions, we can obtain the mass eigenvalue of $a_{-(0)}$ 
defined in Eq.(\ref{def_apm}) as 
\be
 \frac{2m_0^2}{1-\eta}=\frac{1-k^2}{k^2}\Lmd^2. 
\ee
This coincides with the result of Ref.\cite{MSSS}. 

\begin{figure}[t]
 \leavevmode
 \epsfysize=7cm
 \centerline{\epsfbox{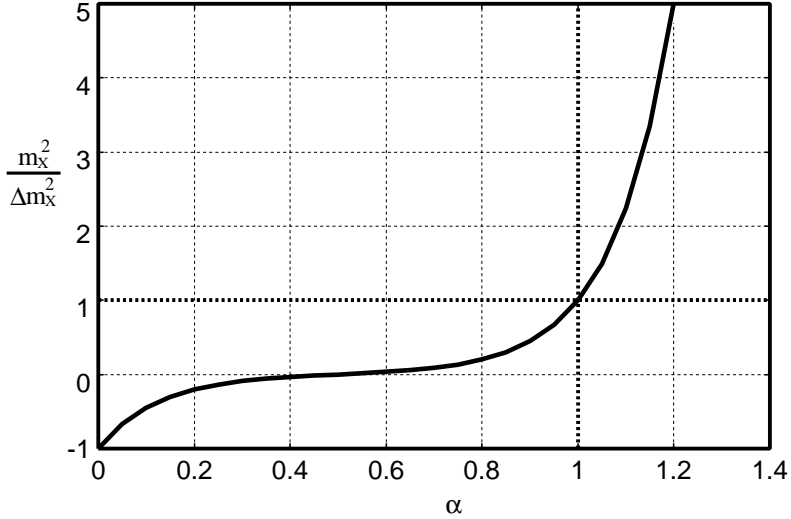}}
 \caption{The ratio~$m_X^2/\Dlt m_X^2$ as a function of $\alpha$.}
 \label{list_rto}
\end{figure}

Fig.\ref{list_rto} shows the ratio~$m_X^2/\Dlt m_X^2$ 
as a function of $\alpha$. 
From this plot, we can see that $m_X^2<|\Dlt m_X^2|$ 
in the region of $0<\alpha<1$. 
This means that a tachyonic mode appears in the effective theory, 
and the classical background~(\ref{bkgd}) is not stable 
in this region. 

In the case of $\alpha>1$, the background is stable 
since $m_X^2>|\Dlt m_X^2|$. 
Here, note that $\alpha$ itself is not the coupling constant 
between $\Phi$ and $X$. 
Indeed, $W_{\rm int}$ in Eq.(\ref{W_mint_exp}) is expanded as  
\be
 W_{\rm int}(\Phi,X)=-\frac{\alpha g}{2}\Phi X^2+\cdots, 
\ee
where ellipsis denotes higher dimensional operators suppressed 
by the scale~$\Lmd$. 
Thus, the true coupling constant is $\alpha g/2$. 
Therefore, the large value of $\alpha$ does not represent the strong coupling 
region as long as $\alpha g/2$ is small. 

In this model, an exact solution~(\ref{exact_sol}) is known. 
In general, however, it is not easy to find such a non-BPS solution 
because we must solve a second order differential equation. 
On the other hand, BPS wall solutions can easily be found 
at least numerically since the BPS equation is a first order 
differential equation. 
As we explained in the previous section, we can calculate 
the parameters of the 3D effective theory with high accuracy 
by using only the BPS wall solutions~$\Acla(x_2)$ and $\Aclb(x_2)$. 
For example, an approximate expression of $\UI(x_2)$ can be obtained 
from Eq.(\ref{apx_UI}) as 
\be
 \UI(x_2)\simeq \left\{\parbox{6cm}{\begin{displaymath}
 \frac{2\Lmd^2}{g}\frac{\cosh(\Lmd x_2)}{\cosh^2(\pi R\Lmd/2)} \;\;\;
 \brkt{-\frac{\pi R}{2}<x_2\leq\frac{\pi R}{2}} \end{displaymath} 
 \begin{displaymath}
 \frac{2\Lmd^2}{g}\frac{1}{\cosh\brc{\Lmd(x_2-\pi R)}} \;\;\;
 \brkt{\frac{\pi R}{2}<x_2\leq\frac{3\pi R}{2}} \end{displaymath}}\right. 
\ee
$\UR(x_2)$ and the mode functions can also be approximated in a similar way.

\section{Interaction with the gauge supermultiplet} \label{gauge_mpt}
In this section, we will derive the 3D effective theory 
including a gauge supermultiplet. 
For simplicity, we will discuss an abelian gauge theory. 
An extension to the non-abelian case is straightforward. 
Let us introduce a vector superfield~$V(x^\mu,\tht,\btht)$ 
in the 4D bulk theory. 
In the Wess-Zumino gauge, its component fields are denoted as 
\be
 V(x^\mu,\tht,\btht)=-\tht\sgm^\mu\btht v_\mu
 +i\tht^2\btht\bar{\lmd}-i\btht^2\tht\lmd+\frac{1}{2}\tht^2\btht^2 D. 
\ee

When we consider interactions with fields 
localized on the BPS wall~$\Acla(x_2)$, it is useful to expand 
$V(x^\mu,\tht,\btht)$ as \cite{sakamura1}
\be
 V(x^\mu,\tht,\btht)
 =e^{-\tht_1\tht_2\der_2}\brc{
 \tht_2\rho^{(1)}(x^m,x_2,\tht_1)+\tht_2^2\sgm^{(1)}(x^m,x_2,\tht_1)}, 
 \label{V_rhsg1}
\ee
where
\bea
 \rho^{(1)}_\alpha (x^m,x_2,\tht_1) \defa 
 -i\brkt{\gm_{(3)}^m\tht_1}_\alpha v_m-\tht_1^2\lmd_{1\alpha}, 
 \nonumber\\
 \sgm^{(1)}(x^m,x_2,\tht_1) \defa 
 -v_2-\tht_1\lmd_2-\frac{1}{2}\tht_1^2 D. 
\eea
Here, 3D Majorana-like spinors~$\lmd_1$ and $\lmd_2$, 
which still have the $x_2$-dependence, are defined 
by the decomposition,\footnote{
The notations of $\lmd_1$ and $\lmd_2$ are different from those 
in Ref.\cite{sakamura1}. 
} 
\be
 \lmd^\alpha = \frac{e^{i\dlt/2}}{\sqrt{2}}(\lmd_2^\alpha+i\lmd_1^\alpha). 
\ee

Using the 4D gauge transformation, we can eliminate the 3D scalar~$v_2$ 
except for its zero-mode~$v_{2(0)}$ \cite{sakamura1}. 

In this case, the 4D superfield 
strength~$W_\alpha\equiv -\frac{1}{4}\bar{D}^2D_\alpha V$ can be written as 
\be
 W_\alpha(y^\mu,\tht)=-\frac{i}{\sqrt{2}}e^{i\dlt/2}
 e^{-\tht_1\tht_2\der_2}e^{-i\tht_2D_1+i\tht_2^2\der_2}
 \brc{u^{(1)}_\alpha(x^m,x_2,\tht_1)+iw^{(1)}_\alpha(x^m,x_2,\tht_1)}, 
\ee
where $u^{(1)}_\alpha$ and $w^{(1)}_\alpha$ are 3D gauge invariant quantities 
and defined as 
\bea
 u^{(1)}_\alpha(x^m,x_2,\tht_1)\defa e^{\tht_1Q_1}\times \lmd_{2\alpha}
 =D_{1\alpha}\sgm^{(1)}+\der_2\rho^{(1)}_\alpha  \nonumber\\
 \eql \lmd_{2\alpha}-\tht_{1\alpha}D +i\brkt{\gm_{(3)}^m\tht_1}_\alpha v_{m2}
 +\tht_1^2\brc{\frac{i}{2}\brkt{\gm_{(3)}^m\der_m\lmd_2}_\alpha
 -\der_2\lmd_{1\alpha}},  \nonumber\\
 w^{(1)}_\alpha(x^m,x_2,\tht_1)\defa e^{\tht_1Q_1}\times \lmd_{1\alpha}
 =\frac{1}{4}D_1^2\rho^{(1)}_\alpha
 +\frac{i}{2}\brkt{\gm_{(3)}^m\der_m\rho^{(1)}}_\alpha \nonumber\\
 \eql \lmd_{1\alpha}+\brkt{\gm_{(3)}^{mn}\tht_1}_\alpha v_{mn}
 -\frac{i}{2}\tht_1^2\brkt{\gm_{(3)}^m\der_m\lmd_1}_\alpha.
\eea
Here, $v_{\mu\nu}\equiv \der_\mu v_\nu-\der_\nu v_\mu$ is a 4D field strength. 

Similarly, when we consider interactions with fields localized 
on the anti-BPS wall~$\Aclb(x_2)$, the following expansion is useful. 
\be
 V(x^\mu,\tht,\btht)
 =e^{\tht_1\tht_2\der_2}\brc{
 -\tht_1\rho^{(2)}(x^m,x_2,\tht_2)+\tht_1^2\sgm^{(2)}(x^m,x_2,\tht_2)},
\ee
where 
\bea
 \rho^{(2)}_\alpha (x^m,x_2,\tht_2) \defa 
 -i\brkt{\gm_{(3)}^m\tht_2}_\alpha v_m-\tht_2^2\lmd_{2\alpha}, 
 \nonumber\\
 \sgm^{(2)}(x^m,x_2,\tht_2) \defa 
 -v_2-\tht_2\lmd_1-\frac{1}{2}\tht_2^2 D. 
\eea 
In this case, $W_\alpha$ can be written as 
\be
 W_\alpha(y^\mu,\tht)=-\frac{i}{\sqrt{2}}e^{i\dlt/2}
 e^{\tht_1\tht_2\der_2}e^{i\tht_1D_2+i\tht_1^2\der_2}
 \brc{u^{(2)}_\alpha(x^m,x_2,\tht_2)-iw^{(2)}_\alpha(x^m,x_2,\tht_2)}, 
\ee
where $u^{(2)}_\alpha$ and $w^{(2)}_\alpha$ are another 3D gauge invariant 
quantities, and defined as 
\bea
 u^{(2)}_\alpha(x^m,x_2,\tht_2)\defa e^{\tht_2Q_2}\times \lmd_{1\alpha}
 =-D_{2\alpha}\sgm^{(2)}-\der_2\rho^{(2)}_\alpha  \nonumber\\
 \eql \lmd_{1\alpha}+\tht_{2\alpha}D -i\brkt{\gm_{(3)}^m\tht_2}_\alpha v_{m2}
 +\tht_2^2\brc{\frac{i}{2}\brkt{\gm_{(3)}^m\der_m\lmd_1}_\alpha
 +\der_2\lmd_{2\alpha}},  \nonumber\\
 w^{(2)}_\alpha(x^m,x_2,\tht_2)\defa e^{\tht_2Q_2}\times \lmd_{2\alpha}
 =\frac{1}{4}D_2^2\rho^{(2)}_\alpha
 +\frac{i}{2}\brkt{\gm_{(3)}^m\der_m\rho^{(2)}}_\alpha \nonumber\\
 \eql \lmd_{2\alpha}+\brkt{\gm_{(3)}^{mn}\tht_2}_\alpha v_{mn}
 -\frac{i}{2}\tht_2^2\brkt{\gm_{(3)}^m\der_m\lmd_2}_\alpha.
\eea
 
The kinetic terms for the gauge supermultiplet can be rewritten 
in terms of $u^{(1)}_\alpha$ and $w^{(1)}_\alpha$, 
{\it i.e.} $\rho^{(1)}_\alpha$ and $\sgm^{(1)}$, as 
\bea
 S^V_{\rm kin} \eql \int\dr^4 x{\rm d}^2\tht\:
 \frac{1}{4}W^\alpha W_\alpha +{\rm h.c.} \nonumber\\
 \eql \int\dr^3 x{\rm d}^2\tht_1\:
 \frac{1}{2}\brc{(u^{(1)})^2-(w^{(1)})^2}.  \label{S_V_kin1}
\eea
Here we have assumed a minimal gauge kinetic function, for simplicity. 

Similarly, Eq.(\ref{S_V_kin1}) can also be rewritten 
in terms of $u^{(2)}_\alpha$ and $w^{(2)}_\alpha$, 
{\it i.e.} $\rho^{(2)}_\alpha$ and $\sgm^{(2)}$, as 
\be
 S^V_{\rm kin} = \int\dr^3 x{\rm d}^2\tht_2\: 
 \frac{1}{2}\brc{(u^{(2)})^2-(w^{(2)})^2}.  \label{S_V_kin2}
\ee

In this case, the mode-expansion of the gauge supermultiplet is trivial. 
That is, 
\bea
 \rho^{(1)\alpha}(x^m,x_2,\tht_1) \eql 
 \frac{1}{\sqrt{2\pi R}}\rho^{(1)\alpha}_{(0)}(x^m,\tht_1) \nonumber\\
 &&+\sum_{n=1}^\infty \frac{1}{\sqrt{\pi R}}
 \brc{\cos\frac{nx_2}{R}\cdot\rho^{(1)\alpha}_{(n+)}(x^m,\tht_1)
 +\sin\frac{nx_2}{R}\cdot\rho^{(1)\alpha}_{(n-)}(x^m,\tht_1)}, \nonumber\\
 \sgm^{(1)}(x^m,x_2,\tht_1) \eql 
 \frac{1}{\sqrt{2\pi R}}\sgm^{(1)}_{(0)}(x^m,\tht_1) \nonumber\\
 &&+\sum_{n=1}^\infty \frac{1}{\sqrt{\pi R}}
 \brc{\cos\frac{nx_2}{R}\cdot\sgm^{(1)}_{(n+)}(x^m,\tht_1) 
 +\sin\frac{nx_2}{R}\cdot\sgm^{(1)}_{(n-)}(x^m,\tht_1)}. \nonumber\\
 \label{rhsg_mode_ex}
\eea
The mode-expansions of $\rho^{(2)\alpha}$ and $\sgm^{(2)}$ 
can be done in a similar way. 

Next, we will consider the gauge interaction with the matter field~$X$. 
In the 4D bulk theory, it is written as 
\be
 S_{\rm matter}^V=\int\dr^4 x{\rm d}^2\tht{\rm d}^2\btht\: \bar{X}e^{2qV}X, 
 \label{Sminimal_cpl}
\ee
where $q$ is a gauge coupling constant. 

Notice that the abelian gauge symmetry should be represented as 
an $SO(2)$ symmetry in our case since 3D superfields are real. 
Thus, the vector superfield~$V$ discussed so far should be understood 
as a $2\times 2$ matrix, 
\be
 V=V_{\rm R}\mtrx{0}{-i}{i}{0}, 
\ee
where $V_{\rm R}$ is a real vector superfield. 
The matter superfield~$X$ should be understood as a 2-component column vector 
whose gauge transformation is 
\be
 X\to \exp\brc{-2q\Lmd\mtrx{0}{-i}{i}{0}}X, 
\ee
where $\Lmd$ is a transformation parameter, which is a chiral superfield. 

Then, by substituting Eqs.(\ref{Phi_vph1}) and (\ref{V_rhsg1}) 
into Eq.(\ref{Sminimal_cpl}) and carrying out the $\tht_2$-integration, 
the following expression can be obtained. 
\bea
 S_{\rm matter}^V \eql \int\dr^3x{\rm d}^2\tht_1{\rm d}x_2\:
 \left\{D_1^\alpha\bar{\vph}^{(1)}_X D_{1\alpha}\vph^{(1)}_X
 +2\Im\brkt{\bar{\vph}^{(1)}_X\der_2\vph^{(1)}_X} 
 +2q\Im\brkt{\bar{\vph}^{(1)}_X\rho^{(1)}D_1\vph^{(1)}_X} \right. 
 \nonumber\\ &&\hspace{2.7cm}\left.
 -\bar{\vph}^{(1)}(2q\sgm^{(1)}-q^2\rho^{(1)2})\vph^{(1)}\right\}. 
\eea
In order to derive the 3D effective theory, 
we should substitute mode-expansions~(\ref{ap_mode-exp2}) 
and (\ref{rhsg_mode_ex}) 
into the above expression. 
Note that there are a lot of light modes in the gauge supermultiplet 
when the compactified radius~$R$ is large compared to the wall width. 
They will remain to be the dynamical degrees 
of freedom after the massive matter modes are integrated out.  

As a result, the interaction terms with the gauge supermultiplet 
can be obtained as follows.  
\bea
 \cL^{(3)}_{\rm int} \eql \int\dr^2\tht_1\:
 \left[q_{(0)}\Im\brkt{{}^t\vph^{(1)}_{X(0)}\rho^{(1)}_{(0)}
 D_1\vph^{(1)}_{X(0)}}+\frac{1}{2}q_{(0)}^2{}^t\vph^{(1)}_{X(0)}
 \rho^{(1)2}_{(0)}\vph^{(1)}_{X(0)}\right. \nonumber\\
 &&\hspace{15mm}
 +\sum_{l=1}^\infty \sum_{s=\pm}q_{(ls)}
 \brc{\Im\brkt{{}^t\vph^{(1)}_{X(0)}\rho^{(1)}_{(ls)}D_1\vph^{(1)}_{X(0)}}
 +q_{(0)}{}^t\vph^{(1)}_{X(0)}\rho^{(1)}_{(ls)}\rho^{(1)}_{(0)}
 \vph^{(1)}_{X(0)}}
 \nonumber\\
 &&\hspace{15mm}
 \left.+\frac{1}{2}\sum_{l,p=1}^\infty\sum_{s,t=\pm} 
 \brkt{q^2}_{(ls,pt)} {}^t\vph^{(1)}_{X(0)}\rho^{(1)}_{(ls)}\rho^{(1)}_{(pt)}
 \vph^{(1)}_{X(0)}\right] \nonumber\\
 &&+\int\dr^2\tht_2\: 
 \left[ q_{(0)}\Im\brkt{{}^t\vph^{(2)}_{X(0)}\rho^{(2)}_{(0)}
 D_2\vph^{(2)}_{X(0)}}+\frac{1}{2}q_{(0)}^2{}^t\vph^{(2)}_{X(0)}
 \rho^{(2)2}_{(0)}\vph^{(2)}_{X(0)} \right. \nonumber\\
 &&\hspace{18mm}
 +\sum_{l=1}^\infty \sum_{s=\pm}q_{(ls)}
 \brc{\Im\brkt{{}^t\vph^{(2)}_{X(0)}\rho^{(2)}_{(ls)}D_2\vph^{(2)}_{X(0)}}
 +q_{(0)}{}^t\vph^{(2)}_{X(0)}\rho^{(2)}_{(ls)}\rho^{(2)}_{(0)}
 \vph^{(2)}_{X(0)}}
 \nonumber\\
 &&\hspace{18mm}
 \left.+\frac{1}{2}\sum_{l,p=1}^\infty\sum_{s,t=\pm} 
 \brkt{q^2}_{(ls,pt)} {}^t\vph^{(2)}_{X(0)}\rho^{(2)}_{(ls)}\rho^{(2)}_{(pt)}
 \vph^{(2)}_{X(0)}\right] \nonumber\\
 &&+\cL^{(3)}_{{\rm SUSY}\hspace{-8mm}
 \begin{picture}(10,5)\line(4,1){23}\end{picture}}\hspace{4mm}, 
\eea
where $q_{(0)}\equiv q/\sqrt{2\pi R}$ is the 3D gauge coupling constant, 
and the other effective couplings are defined as 
\bea
 q_{(n+)}\defa \frac{g}{\sqrt{\pi R}}\int\dr x_2 \:
 \uRX{0}^2\cos\frac{nx_2}{R}
 =\frac{g}{\sqrt{\pi R}}\int\dr x_2\:
 \uIX{0}^2\cos\frac{n(x_2-\pi R)}{R}, \nonumber\\
 q_{(n-)}\defa \frac{g}{\sqrt{\pi R}}\int\dr x_2 \:
 \uRX{0}^2\sin\frac{nx_2}{R}
 =\frac{g}{\sqrt{\pi R}}\int\dr x_2\:
 \uIX{0}^2\sin\frac{n(x_2-\pi R)}{R}, \nonumber\\
 \brkt{q^2}_{(n+,m+)}\defa \frac{g^2}{\pi R}\int\dr x_2\:
 \uRX{0}^2\cos\frac{nx_2}{R}\cos\frac{mx_2}{R},  \nonumber\\
 \brkt{q^2}_{(n+,m-)}\defa \frac{g^2}{\pi R}\int\dr x_2\:
 \uRX{0}^2\cos\frac{nx_2}{R}\sin\frac{mx_2}{R},  \nonumber\\
 &\vdots&
\eea
and $\cL^{(3)}_{{\rm SUSY}\hspace{-8mm}
\begin{picture}(10,5)\line(4,1){23}\end{picture}}$\hspace{4mm} 
denotes the SUSY breaking terms as follows. 
\bea
 \cL^{(3)}_{{\rm SUSY}\hspace{-8mm}
 \begin{picture}(10,5)\line(4,1){23}\end{picture}}\hspace{4mm}
 \eql -\eta_X{}^t\!\brkt{\der^ma^{(1)}_{X(0)}+iq_{(0)}v_{(0)}^ma^{(1)}_{X(0)}}
 \brkt{\der_ma^{(2)}_{X(0)}+iq_{(0)}v_{(0)m}a^{(2)}_{X(0)}} 
 \nonumber\\
 &&+i\eta_X q_{(0)}\brkt{a^{(1)}_{X(0)}\lmd_{1(0)}\psIX{0}
 +a^{(2)}_{X(0)}\lmd_{2(0)}\psRX{0}}
 -i\eta_X q_{(0)}\psRX{0}v_{2(0)}\psIX{0} \nonumber\\
 &&+\cdots, 
\eea
where ellipsis denotes terms involving the Kaluza-Klein modes for 
the gauge supermultiplet. 

The gaugino mass terms are induced at loop-level in this model.

\section{Summary and discussion} \label{summary}
We have provided a useful method for the derivation 
of the effective theory in a system 
where BPS and anti-BPS domain walls coexist. 
Although the corresponding field configuration~$\Acl(x_2)$ is non-BPS, 
it can approximately be regarded as a BPS wall or an anti-BPS wall  
in the neighborhood of each domain wall. 
So the light modes localized on one of the walls 
form (approximate) supermultiplets for the corresponding 3D $\cN=1$ SUSY. 
Due to the existence of the other wall, however, 
such approximate SUSYs are broken. 
Thus, the 3D effective theory for this system consists of two parts; 
a supersymmetric part described by 3D superfields and 
a SUSY breaking part. 
All parameters of the effective theory can be calculated systematically 
in our method. 


The SUSY breaking terms can be classified into two types. 
The first-type ones involve only modes localized on the same wall. 
Such terms are induced due to the deviation of the background 
from the BPS wall, 
which is mainly characterized by $\UR(x_2)$ or $\UI(x_2)$ 
in Eq.(\ref{U_in_vpcl}). 
Indeed, all parameters for such terms  
are expressed by overlap integrals involving $\UR(x_2)$ or $\UI(x_2)$. 
The second-type ones are 
direct couplings between modes localized on the different walls. 
Both types of the SUSY breaking terms receive 
an exponential suppression in terms of 
the ratio of the wall distance~$\pi R$ to the wall width~$\lmd_w$.\footnote{
In the model discussed in Sect.\ref{exmpl_model}, 
$\lmd_w=k/\Lmd\simeq 1/\Lmd$. }
In particular, all SUSY breaking terms vanish at tree-level 
in the ``thin-wall'' limit, {\it i.e.} $\pi R/\lmd_w\to\infty$. 
This situation corresponds to the {\it pseudo-supersymmetry} 
discussed in Ref.\cite{klein}.  
In this limit, SUSY breaking is induced through the loop effects 
involving the bulk fields, such as the gauge field 
discussed in Sect.\ref{gauge_mpt}. 
In the case that $\pi R/\lmd_w$ is finite, on the other hand, 
the tree-level terms of SUSY breaking discussed in this paper arise 
besides the loop-induced ones. 
It depends on the ratio~$\pi R/\lmd_w$ whether the tree-level ones dominate 
over the loop-induced ones or not. 

From the SUSY breaking mass terms calculated in our method, we have shown that 
maximal mixings occur between the bosonic zero-modes localized 
on the different walls, while the fermionic zero-modes remain to be localized 
on each wall. 
This means that oscillations will occur between the bosonic modes localized 
on the different walls\footnote{
This is a similar situation to the one in Ref.\cite{hofmann} 
where the multi-BPS-wall configurations are considered.}. 

An effective theory on the domain wall can also be derived 
by using the nonlinear realization technique for the space-time symmetries. 
In particular, we can obtain an effective theory on a BPS wall 
by regarding the existence of the wall 
as the partial SUSY breaking \cite{bagger}. 
This method is powerful for the general discussion of the BPS walls 
because it uses only information on symmetries. 
The relation between this method and the mode-expansion method 
\cite{DS,sakamura1} is discussed in Ref.\cite{sakamura2}. 
Similarly, we can derive an effective theory on a {\it non-BPS} wall 
by the nonlinear realization method \cite{nitta}. 
Since the wall-antiwall system is non-BPS, 
we can apply this method to our case. 
However, the nonlinear realization method cannot respect a specific structure 
of the configuration, such as the wall-antiwall structure. 
Thus, in order to discuss the specific features of the wall-antiwall system, 
we have to take the mode-expansion method as we have done in this paper.  
If you would like to know the relation between our result and 
the result of Ref.\cite{nitta}, 
you should combine our method with the result of Ref.\cite{sakamura2}. 

In this paper, we have concentrated ourselves on the case 
that the K\"{a}hler potential is minimal, for simplicity. 
An extension to the non-minimal K\"{a}hler case is possible, 
although the classical background~$\Acl(x_2)$ and the mode functions 
cannot be obtained analytically in that case. 
Thus, the BPS-wall approximation discussed in Sect.\ref{BPSapx} 
is useful in such a case because solving the BPS equations 
(the first order differential equations) is much easier than 
solving the equations of motion (the second order differential equations) 
in the numerical calculation. 

When we introduce the matter fields, we must take care of 
the strength of the couplings between the matter and the wall fields. 
As we have demonstrated in Sect.\ref{exmpl_model}, 
a weak coupling between them leads to a tachyonic scalar mode 
in the effective theory. 
This means that the background field configuration is unstable 
unless the matter modes are strongly localized on the walls. 
Roughly speaking, if the matter modes are localized 
within the wall width~$\lmd_w$, the background is stable. 
Therefore, the fat brane scenario \cite{arkani,kaplan} is allowed 
in the wall-antiwall system. 
Using the above tachyonic mode as the ``Higgs'' field, 
it might be possible to propose a new mechanism of symmetry breaking. 

In the superstring theory, it is well-known that 
BPS D-branes can be described by the kink solutions 
in the tachyon field theory \cite{MZ}. 
In particular, a system where D$p$ brane and anti-D$p$ brane coexist 
is represented by a wall-antiwall solution 
in an effective field theory on a non-BPS D$(p+1)$-brane \cite{hashimoto}. 
Thus, our method discussed here is useful 
in the discussion of the tachyon condensation \cite{tachyon} 
in the superstring theory. 

In order to construct a more realistic model, we have to investigate 
a wall-antiwall system in a 5D SUSY theory. 
However, 5D SUSY theories are difficult to handle 
because they have many SUSY (at least eight supercharges).  
In fact, 5D SUSY theories are required to be nonlinear sigma models 
in order to have a BPS domain wall \cite{nsakai}, 
due to their restrictive forms of the scalar potential. 
However, our approach suggests how to derive the 4D effective theory 
for the wall-antiwall system, 
once we find a method for deriving an effective theory on a BPS wall 
in five dimensions.  
Expanding the discussion to the supergravity is also an interesting 
subject.

\vspace{5mm}

\begin{center}
{\bf Acknowledgments}
\end{center}
The author thanks Nobuhito Maru, Norisuke Sakai and Ryo Sugisaka 
for a collaboration of previous works which motivate this work. 
The author also thanks the Yukawa Institute for Theoretical Physics 
at Kyoto University, where this work was initiated during 
the YITP-W-02-04 on ``Quantum Field Theory 2002''.

\appendix
\section{Notations} 
Basically, we follow the notations of Ref.\cite{WB} 
for the 4D bulk theory. 
\subsection{Notations for 3D theories} 
The notations for the 3D theories are as follows. 

We take the space-time metric as 
\be
 \eta^{mn}=\diag(-1,+1,+1). 
\ee

The 3D $\gm$-matrices, $(\gm_{(3)}^m)_\alpha^{\;\beta}$, 
can be written by the Pauli matrices as 
\be
 \gm_{(3)}^0=\sgm^2, \;\;\;
 \gm_{(3)}^1=-i\sgm^3, \;\;\;
 \gm_{(3)}^3=i\sgm^1, 
\ee
and these satisfy the 3D Clifford algebra, 
\be
 \brc{\gm_{(3)}^m,\gm_{(3)}^n}=-2\eta^{mn}. 
\ee

The spinor indices are raised and lowered by multiplying $\sgm^2$ 
from the left. 
\be
 \psi_\alpha=\brkt{\sgm^2}_{\alpha\beta}\psi^\beta, \;\;\;
 \psi^\alpha=\brkt{\sgm^2}^{\alpha\beta}\psi_\beta. 
\ee
We take the following convention of the contraction of spinor indices. 
\be
 \psi_1\psi_2\equiv\psi_1^\alpha\psi_{2\alpha}
 =\brkt{\sgm^2}_{\alpha\beta}\psi_1^\alpha\psi_2^\beta=\psi_2\psi_1. 
\ee

\section{Properties of $\Acl(x_2)$} \label{Acl_prop}
Here, we collect some properties of the classical 
configuration~$\Acl(x_2)$. 

The equation of motion for the classical solution~$\Acl(x_2)$ is  
\be
 \der_2^2\Acl-W'(\Acl)\bar{W}''(\bar{A}_{\rm cl})=0. 
\ee
By multiplying $\der_2\bar{A}_{\rm cl}(x_2)$ with this equation, 
we will obtain 
\be
 \frac{1}{2}\der_2\brc{|\der_2\Acl|^2-|W'(\Acl)|^2}=0. 
\ee
This means that 
\be
 |\der_2\Acl|^2-|W'(\Acl)|^2=c,  \label{def_c}
\ee
where $c$ is a real constant. 
In the case that $\Acl(x_2)$ is a BPS wall configuration, 
we can see that $c=0$ from the BPS equation. 
For the wall-antiwall configuration, $c$ is a tiny constant. 
Typically, it is exponentially suppressed by the wall distance~$\pi R$. 
In the model discussed in Sect.\ref{exmpl_model}, for instance, 
\be
 c=\brkt{\frac{\Lmd^2}{gk}}^2(1-k^2)
 \simeq 16\frac{\Lmd^4}{g^2}e^{-\pi R\Lmd}. 
\ee
In particular, in the case of the real configuration 
mainly discussed in this paper, Eq.(\ref{def_c}) can be rewritten as 
\be
 \UR(x_2)\UI(x_2)=V_0\uR{0}(x_2)\uI{0}(x_2)=c. \label{UR-UI_prdct} 
\ee
Thus, $c$ is related to $\eta$ in Eq.(\ref{L_quad}) through 
\be
 c=\frac{V_0}{2\pi R}\eta.  \label{c_expl}
\ee

Under the assumption~(\ref{W_assumption}), 
the following relations hold. 
\bea
 W''(\Acl(x_2-\pi R)) \eql -W''(\Acl(x_2)), \nonumber\\
 W''(\Acl(-x_2)) \eql W''(\Acl(x_2)),  \label{Wpp_prop}
\eea
\be
W'''(\Acl(x_2-\pi R))=\mp W'''(\Acl(x_2)). \label{Wppp}
\ee
The upper sign in R.H.S. of Eq.(\ref{Wppp}) corresponds to the case that 
the configuration~$\Acl(x_2)$ is stable, 
and the lower sign corresponds to the case of an unstable configuration, 
respectively. 

Thus, together with Eq.(\ref{piR_translation}), 
we can obtain the following relation. 
\be
 \sbk{W'''(\Acl)\UI}(x_2-\pi R) 
 =-\sbk{W'''(\Acl)\UR}(x_2).  \label{WU_piRtrslt}
\ee

\end{document}